\documentclass{aa}

\usepackage{graphics}

\newcounter{refcts}

\begin{document}

\title{
Coronal density diagnostics with Helium-like triplets:\\
CHANDRA--LETGS observations of Algol, Capella, Procyon, $\epsilon$~Eri,
$\alpha$~Cen A\&B, UX~Ari, AD~Leo, YY~Gem, and HR\,1099}

 \author{J.-U.~Ness\inst{1}
        \and
        J.H.M.M.~Schmitt\inst{1}
        \and
        V.~Burwitz\inst{2}
        \and
        R.~Mewe\inst{3}
        \and
	A.J.J.~Raassen\inst{3,4}
	\and
	R.L.J.~van der Meer\inst{3}
	\and
	P.~Predehl\inst{2}
	\and
	A.C.~Brinkman\inst{3}
}

\institute{
 Universit\"at Hamburg, Gojenbergsweg 112, D-21029 Hamburg, Germany
 \and
  Max-Planck-Institut f\"ur Extraterrestrische Physik (MPE), Postfach 1603,
  D-85740 Garching, Germany
 \and
 Space Research Organization Netherlands (SRON),
 Sorbonnelaan 2, 3584 CA Utrecht, The Netherlands
  \and
 Astronomical Institute "Anton Pannekoek", Kruislaan 403,
 1098 SJ Amsterdam, The Netherlands}

\authorrunning{J.U. Ness, J.H.M.M. Schmitt et al.}
\titlerunning{Coronal density diagnostics with Helium-like triplets}

\offprints{J.-U.\ Ness}
\mail{jness@hs.uni-hamburg.de}
\date{received April 24, 2002; accepted August 2, 2002}

\abstract{
We present an analysis of ten cool stars (Algol, Capella, Procyon,
$\epsilon$~Eri, $\alpha$~Cen A\&B, UX~Ari, AD~Leo, YY~Gem, and HR\,1099)
observed with the Low Energy
Transmission Grating Spectrometer (LETGS) on board the {\it Chandra} X-ray
Observatory. This sample contains all cools stars observed with the LETGS
presently available to us with integration times sufficiently long to warrant
a meaningful spectral analysis. Our sample comprises inactive, moderately
active, and hyperactive stars and samples the
bulk part of activity levels encountered in coronal X-ray sources. We use the
LETGS spectra to carry out density and temperature diagnostics with an emphasis
on the H-like and the He-like ions. We find a correlation between line flux
ratios of the
Ly$_\alpha$ and He-like resonance lines with the mean X-ray surface flux. We
determine densities using the He-like triplets. For active stars we find no
significant deviations from the low-density limit for the ions of Ne, Mg, and
Si, while the measured line ratios for
the ions of C, N, and O do show evidence for departures from the low-density
limit in the active stars, but not in the inactive stars. Best measurements can
be made for the O\,{\sc vii} triplet where we find significant deviations from
the low-density limit for the stars Algol, Procyon, YY~Gem, $\epsilon$~Eri, and
HR\,1099. We discuss the influence of radiation fields on the
interpretation of the He-like triplet line ratios in the low-Z ions, which is
relevant for Algol, and the influence of dielectronic satellite lines, which is
relevant for Procyon. For the active stars YY~Gem, $\epsilon$~Eri, and HR\,1099
the low f/i ratios can unambiguously be attributed to high densties in the range
1--3\,10$^{10}$\,cm$^{-3}$ at O\,{\sc vii} temperatures.
We find our LETGS spectra to be an
extremely useful tool for plasma diagnostics of stellar coronae.
}

\maketitle

\section{Introduction}
\label{intro}

The coronal X-ray emission from the Sun is spatially correlated with
photospheric regions exhibiting magnetic field concentrations. Therefore
spatially unresolved X-ray emission from other solar-like stars is
commonly used as a tracer for stellar magnetic activity. The specific
advantage of X-ray measurements is that any stellar X-ray emission
exclusively comes from the corona, and unlike other activity tracers, is not
affected by photospheric emissions, rapid rotation, turbulent broadening etc.
X-ray observations of stars carried out with the {\it Einstein Observatory}
(cf. Vaiana et al. \cite{vai81}) and with ROSAT (Schmitt \cite{schm97})
revealed the ubiquity of stellar X-ray emission among stars placed in the
Hertzsprung-Russell diagram. The extensive ROSAT surveys showed that
essentially all late-type solar-like stars with outer convection are
surrounded by hot (T $\ge$ 1\,MK) coronae (Schmitt \cite{schm97}).
The X-ray luminosity of a given type of star can vary over
four orders of magnitude over the sample, and rotation appears to be
the most important parameter characterizing the level of X-ray emission of
cool stars (cf. Pallavicini et al. \cite{pal81}). The solar corona as
observed with modern X-ray and XUV telescopes on board YOHKOH, SOHO,
or TRACE is found to be extremely structured, and even in the high angular
resolution TRACE images there appears to be spatially unresolved fine
structure. Yet spatially resolved X-ray observations of stellar coronae are
currently not feasible. Information on the spatial structure of stellar
coronae can in principle be derived from X-ray light curves of suitably
chosen stars such as eclipsing binaries, yet the
actual information derivable from such data is rather limited; a
discussion of the pre-XMM and pre-{\it Chandra} results together with the
difficulties and limitations
of light curve inversions is given by Schmitt (\cite{schm98}).

How do stars manage to produce far more X-ray output than the Sun?
One of the basic assumptions used in the interpretation of stellar
X-ray emission is that the building blocks, the solar corona is composed
of, are also those that make up the X-ray emission of stars. This assumption
can only be tested by spectroscopic investigations. The {\it Einstein
Observatory} and ROSAT provided data with rather modest spectral
resolution. A few measurements of selected stars were carried out with the
transmission gratings available on board of the {\it Einstein} and EXOSAT
satellites. Higher spectral resolution information, albeit only at wavelengths
longward of 90\,\AA, but not in the X-ray range, was provided by the EUVE
spectrometers. The low resolution proportional counter spectra obtained
with the {\it Einstein} and ROSAT satellites were fitted with plasma emission
models in order to derive plasma temperatures and emission measures (see, for
example, Schmitt et al. \cite{schm90}); later also elemental abundances were
included as fit parameters (see, for example, Antunes et al. \cite{ant94}).
However, the previous measurements did not allow to measure densities $n_e$ and
emission measures $EM$ independently, such that no emitting volumes $V$ could
be derived from $EM= n_e^2V$. Therefore no information about loop sizes was
accessible. With the new high resolution spectra obtained with the {\it Chandra}
and XMM-Newton grating spectrometers it is possible for the first time to
measure individual lines in the X-ray range with reasonable effective areas
over a wide bandpass for a larger sample of stars in the same fashion as X-ray
emission lines from the solar corona have been obtained and analyzed for many
years (e.g., Doyle \cite{doyle80}, McKenzie \& Landecker \cite{mck82}, Gabriel
et al. \cite{gbf88}). Specifically, He-like triplets can be observed
for a variety of elements (carbon, nitrogen, oxygen, neon, magnesium, and
silicon) and stars. The theory of He-like triplets can be tested in
active and inactive stars and density information can be derived; this
information can be supplemented by other line ratios, e.g.,
Fe\,{\sc xxi} lines (measured with the LETGS), which also yield
density constraints. The density-sensitive line ratios
are the missing link relating emission measure and
volumes. Further, emission measure weighted ``effective'' temperatures can
be derived from suitable line ratios, in particular the ratios between the
Ly$_{\alpha}$ and He-like resonance lines.

First results from XMM-RGS measurements were presented by, e.g., Audard et al.
(\cite{aud1}, \cite{aud2}) and G\"udel et al. (\cite{gued1}, \cite{gued2})
with special focus on global modeling techniques to derive temperature
distributions and abundances for the Castor system, Capella, AB~Dor, and
HR\,1099.
Measurements with the MEG on board {\it Chandra} were presented by, e.g.,
Canizares et al. (\cite{caniz00}), Ayres et al. (\cite{ayres01}), and Phillips
et al. (\cite{phil01}). The MEG and HEG provide the highest resolution at
energies $>$ 1 keV and Brickhouse et al. (\cite{brick01}) report the
measurements of orbit related line shifts in 44~Boo, thus
eventually opening up the road to X-Ray Doppler imaging of stellar coronae.
Plasma diagnostics carried out with LETGS spectra were presented by, e.g., Mewe
et al. (\cite{mewe_cap}) on Capella, Ness et al. (\cite{ness_cap}) on Capella
and Procyon, and Ness et al. (\cite{ness_alg}) on Algol; these authors use
line ratios in order to derive plasma densities and temperatures.\\
The purpose of this paper is to summarize results from recent LETGS
measurements of a set of ten cool stars with a special focus on the
density diagnostics with He-like triplets. Albeit some results on He-like
triplets were published earlier (e.g., Ness et al. \cite{ness_cap},
Mewe et al. \cite{mewe_cap}, Ness et al. \cite{ness_alg}, Stelzer et al.
\cite{stelz02}, Raassen et al. \cite{raa}, Audard et al. \cite{aud3}), we will
use these already published results for comparison with our new results
on UX~Ari, $\epsilon$~Eri, $\alpha$~Cen A and B, and AD~Leo, which are
presented here for the first time.
\\
The specific information that can be derived from He-like line ratios is the
plasma density. The theory of such triplets is described extensively in the
literature; we refer to Gabriel \& Jordan (\cite{gj69}), who developed the
theory, and to Blumenthal et al. (\cite{bl72}), Mewe \& Schrijver (\cite{ms78}),
Pradhan et al. (\cite{prad81}), Pradhan \& Shull (\cite{ps81}), and recently
Porquet et al. (\cite{por01}), who refined and revised the theory. In this
paper we use the relation
\begin{equation}
\frac{\rm f}{\rm i} \ = \frac {R_{\rm 0}} {1 + \phi/\phi_c + n_e/N_c}\,,
\end{equation}
denoting the forbidden line with f and the intercombination line with i
with the low density limit $R_{\rm 0}$, the radiation term $\phi/\phi_c$ (the
values derived in Sect.~\ref{radfield} are listed in Tables~\ref{restab1} to
\ref{restab3}), and the electron density $n_e$; $N_c$ is the so-called
critical density, which leads to an f/i-ratio just in between the high- and
low-density limits. Thus, given an X-ray measurement of the f/i-ratio and
knowledge of $R_{\rm 0}$, $N_c$, and $\phi/\phi_c$ from other sources, the
plasma electron density $n_e$ can be inferred.\\
Our paper will be structured as follows: We first give a short description of
the instrument with its capabilities and provide an overview of our sample of
stars. We then introduce the new spectra and give a detailed description of
the analysis presenting the measured line counts and describing the methods
applied to analyze the obtained results. Our results of total X-ray
luminosities, temperatures, and He-like densities
for the ions Si\,{\sc xiii}, Mg\,{\sc xi}, Ne\,{\sc ix}, O\,{\sc vii},
N\,{\sc vi}, and C\,{\sc v} will be presented in Sect.~\ref{result}.
Our conclusions are given in Sect.~\ref{conclusio}.

\section{The instrument}
\label{instrument}

The new generation of X-ray telescopes and X-ray spectrometers
on board {\it Chandra} and XMM-Newton
has opened the world of spectroscopy to X-ray astronomy with specific emphasis
on sensitivity (XMM-RGS) and spectral resolution ({\it Chandra} LETGS and
HETGS). Spectroscopic measurements can be performed with both instruments using
the Reflection Grating Spectrometer (RGS) on board XMM and the High Energy
Grating (HEG), Medium Energy Grating (MEG), and the Low Energy
Transmission Grating (LETGS) on board {\it Chandra}. While RGS, HEG, and MEG
provide large effective area for high energies in the range 5 to 40\,\AA\ with
high resolution, the LETGS covers a much larger wavelength range from 5 --
175\,\AA\ with high spectral resolution encompassing both the ROSAT bandpass
(5 -- 124\,\AA) and the {\it Einstein} bandpass (3 -- 84\,\AA). As a result of
this large band pass X-ray fluxes and X-ray luminosities corresponding to ROSAT
or {\it Einstein} band passes can be measured from the LETGS data without the
use of any plasma emission model by simply integrating the photon energies
over all wavelength bins covering the corresponding wavelength ranges (cf.
Sect.~\ref{spectra}); the only corrections to be applied are those from
interstellar absorption which are rather small for nearby stars. We further note
that in the long wavelength region the spectral resolution of the LETGS even
exceeds the spectral resolutions of the other instruments. One specific
advantage of the LETGS resulting from its large wavelength range covered is the
fact that the O\,{\sc vii} triplet at around 22\,\AA, the N\,{\sc vi} triplet
at around 29\,\AA, and the C\,{\sc v} triplet at around 41\,\AA\ can all be
measured in one
spectrum. The O\,{\sc vii} triplet is covered by all instruments, but the MEG
effective area is already small and steeply falling over the O\,{\sc vii}
triplet region. The N\,{\sc vi} triplet is covered by the XMM-RGS
and the LETGS, the C\,{\sc v} triplet is only covered by the LETGS as well
as the long wavelengths $>$100\,\AA\ which contain a number of lines from
highly ionized Fe~ions. This range was used by, e.g., Mewe et al.
(\cite{mewe_cap}) and Ness et al. (\cite{ness_alg}) for an independent density
analysis with Fe\,{\sc xxi} line ratios (cf. Mason et al. \cite{mason84}).\\

\section{The sample of stars}
\label{stars}

\begin{table*}
\caption[ ]{\label{sprop}Summary of stellar properties and measurement of
X-ray luminosities for the stars. Effective areas are taken from Pease et
et al. (\cite{dpease00}). ROSAT measurements are taken from, e.g., H\"unsch
et al. (\cite{huen}, \cite{huen98}). References for stellar radii and distances
are given with superscripts.}
\renewcommand{\arraystretch}{1.2}
{\scriptsize
\begin{tabular}{l|cccccccccc}
\hline
&Algol&Capella&Procyon&$\epsilon$~Eri& \multicolumn{2}{c}{$\alpha$~Cen}&UX~Ari&AD~Leo&YY~Gem&HR\,1099\\
&&&&&A&B&&\\
\hline
HD&19356&34029&61421&22049&128620&128621&21242&Gl388&60179C&22468\\
Spectr. Type&K2IV&G1III+G8&F5 IV-V&K2V&G2V&K0V&G5V/K0IV&dM4.5Ve&dMIe/dMIe&K1IV/G5IV\\
&&\ \ /K0III&&&&&&&&\\
d/pc&28$^{\tiny [\cite{pery}]}$&13$^{\tiny [\cite{stra}]}$&3.5&3.22&\multicolumn{2}{c}{1.34}&50$^{\tiny [\cite{stra}]}$&4.9&14.7&28.97\\
%d/pc&28&13&3.5&3.22&\multicolumn{2}{c}{1.34}&50&4.9&14.7&28.97\\
%$R_\star/R_\odot$&3.5&9.2/12.2&2.06&0.81&1.23&0.80&0.93/$>$4.7&0.5&0.66/0.58&3.9/1.3\\
$R_\star/R_\odot$&3.5$^{\tiny [\cite{ric}]}$&9.2$^{\tiny [\cite{hum}]}$&2.06$^{\tiny [\cite{irw}]}$&0.81$^{\tiny [\cite{noy}]}$&1.23$^{\tiny [\cite{sod}]}$&0.80$^{\tiny [\cite{sod}]}$&0.93/$>$4.7$^{\tiny [\cite{stra}]}$&0.5&0.66/0.58&3.9/1.3\\
\hline
t$_{exp}$/ksec&81.4&218.5&140.7&108.0&\multicolumn{2}{c}{81.5}&112.76&48.5&59&97.5\\
$^aL_X$ [10$^{28}$\,erg/s]&1444&255&2.43&20.9&0.65& 0.52&2302&6.7&54.4&1585.6\\
$^aF_X$ [10$^5$\,erg/cm$^2$/s]&191.4&4.89&0.93&51.73&0.70&1.32&169.24&43.52&245.4&169.3\\
$^bL_X$[10$^{28}$\,erg/s]&1101&239&1.64&2.22&0.30&0.26&1539&5.7&46.8&1156.2\\
ROSAT [10$^{28}$\,erg/s]&661&419.16&1.9&2.1&0.13$^{d}$&0.20$^{d}$&1205&7.22&82.37&1512.1\\
$^cL_X$[10$^{28}$\,erg/s]&1262&219&0.86&1.85&0.17&0.14&1716&5.0&45.7&1273.8\\
{\it Einstein} [10$^{28}$\,erg/s]&371&281&1.21&1.47&0.12$^{e}$&0.28$^{e}$&2900&8.32&21.5&700\\
\hline
\multicolumn{4}{l}{$^a$Range 0.07-2.5\,keV (LETGS range: 5-175\,\AA)}&&&&&&&\\
\multicolumn{4}{l}{$^b$Range 0.1-2.4\,keV (ROSAT range: 5.2-124\,\AA)}&&&&&&&\\
\multicolumn{4}{l}{$^c$Range 0.15-4\,keV ({\it Einstein} range: 3-83\,\AA)}&&&&&&&\\
\multicolumn{4}{l}{$^d$Schmitt et al. 1998, ASP Conf., 154, 463}&&&&&&&\\
\multicolumn{4}{l}{$^e$Golub et al. 1982, ApJ, 253, 242}&&&&&&&\\
\end{tabular}\\
\vspace{-1.9cm}
\begin{list}{}{\setlength{\itemindent}{5.8cm}\setlength{\itemsep}{-.05cm}}
 \setcounter{refcts}{1}
 \bibitem[\therefcts]{hum} $^{[\therefcts]}$Hummel, C.A., Armstrong, J.T., Quirrenbach, A., et al. 1994, ApJ, 107(5), 1859
\stepcounter{refcts}
 \bibitem[\therefcts]{irw} $^{[\therefcts]}$Irwin, A.W., Fletcher, J.M., Yang, S.L.S., et al. 1992, PASP, 104, 489
\stepcounter{refcts}
 \bibitem[\therefcts]{noy} $^{[\therefcts]}$Noyes, R. W., Baliunas, S. L., Belserene, E., et al. 1984, ApJ, 285, L23
\stepcounter{refcts}
 \bibitem[\therefcts]{pery} $^{[\therefcts]}$Perryman, M.A.C., Lindegren, L., Kovalevsky, J., et al. 1997, A\&A, 323L, 49
\stepcounter{refcts}
 \bibitem[\therefcts]{ric} $^{[\therefcts]}$Richards, M.T. 1993, ApJ, 86, 255
\stepcounter{refcts}
 \bibitem[\therefcts]{sod} $^{[\therefcts]}$Soderblom 1986, A\&A, 158, 273
\stepcounter{refcts}
 \bibitem[\therefcts]{stra} $^{[\therefcts]}$Strassmeier, K. G., Hall, D. S.,
 Fekel, F. C., et al. 1993, A\&A, 100, 173
\end{list}
}
%\end{tabular}
\vspace{-.5cm}
\begin{flushleft}
\renewcommand{\arraystretch}{1}
\end{flushleft}
\end{table*}

The properties of our sample stars are summarized in Table~\ref{sprop};
we also provide previously measured X-ray luminosities
as well as the X-ray luminosities derived from our LETGS observations. As is
clear from Table~\ref{sprop}, our sample of stars comprises a wide
range of X-ray
luminosities $L_X$ from few $10^{27}$ up to $10^{32}$\,erg/sec, which in fact
covers almost the whole range of X-ray luminosities encountered for
cool stars. The sample consists of three rather inactive dwarf stars in the
immediate solar neighborhood, i.e., $\alpha$~Cen A and B and Procyon, of the
active late-type dwarf stars AD~Leo, YY~Gem, and $\epsilon$~Eri, and of the
active binary systems Algol, Capella, HR\,1099, and UX~Ari, where rapidly
rotating evolved stars are responsible for the observed activity. We note in
passing that the
LETGS spectrally resolved $\alpha$~Cen A and B for the first time. A detailed
analysis of this spectrum will be presented by Raassen et al.~\cite{raacen}.

\section{Observations and data analysis}

\subsection{Data extraction and reference to previous work}

All the LETGS datasets were processed using the standard pipeline processing.
The extraction of the spectra from the microchannel plate and the determination
of the instrumental background are preformed in the same fashion as described
by Ness et al. (\cite{ness_cap}). The right and left dispersed spectra were
co-added in order to increase the signal-to-noise ratio; this procedure is
justified by Ness et al. (\cite{ness_cap}) using Gaussian profiles. The number
of counts in the individual lines is obtained with the CORA program
(Ness \& Wichmann \cite{new02}) developed
and described by Ness et al. (\cite{ness_cap}). The effective areas used for
converting counts into fluxes were provided by Deron Pease (Oct. 2000) and all
fluxes derived in this paper are based on this calibration.\\
In this paper we use the results obtained for Procyon and Capella by Ness et
al. (\cite{ness_cap}) for oxygen, nitrogen, and carbon, but corrected for new
effective areas. The higher ionization stages for Capella have been analyzed by
Mewe et al. (\cite{mewe_cap}), but we re-analyze these ions using the larger
set of available observations as listed in Ness et al. (\cite{ness_cap}) and
revising the results with the more recent effective areas; this reanalysis also
ensures the maximally possible uniformity in data reduction. The results for
Algol are taken from Ness et al. (\cite{ness_alg}) without any further
correction except for Ne\,{\sc ix}.
The analysis of the Ne\,{\sc ix} triplet is new for all the stars except
for Algol. In this paper we use a different approach and apply this procedure
to Algol as well. We list all H-like and He-like results for the
purpose of comparison in Tables~\ref{flux_1} and \ref{flux_2}.\\

\subsection{The spectra}
\label{spectra}

\begin{figure*}[!ht]
 \resizebox{\hsize}{!}{\rotatebox{90}{\includegraphics{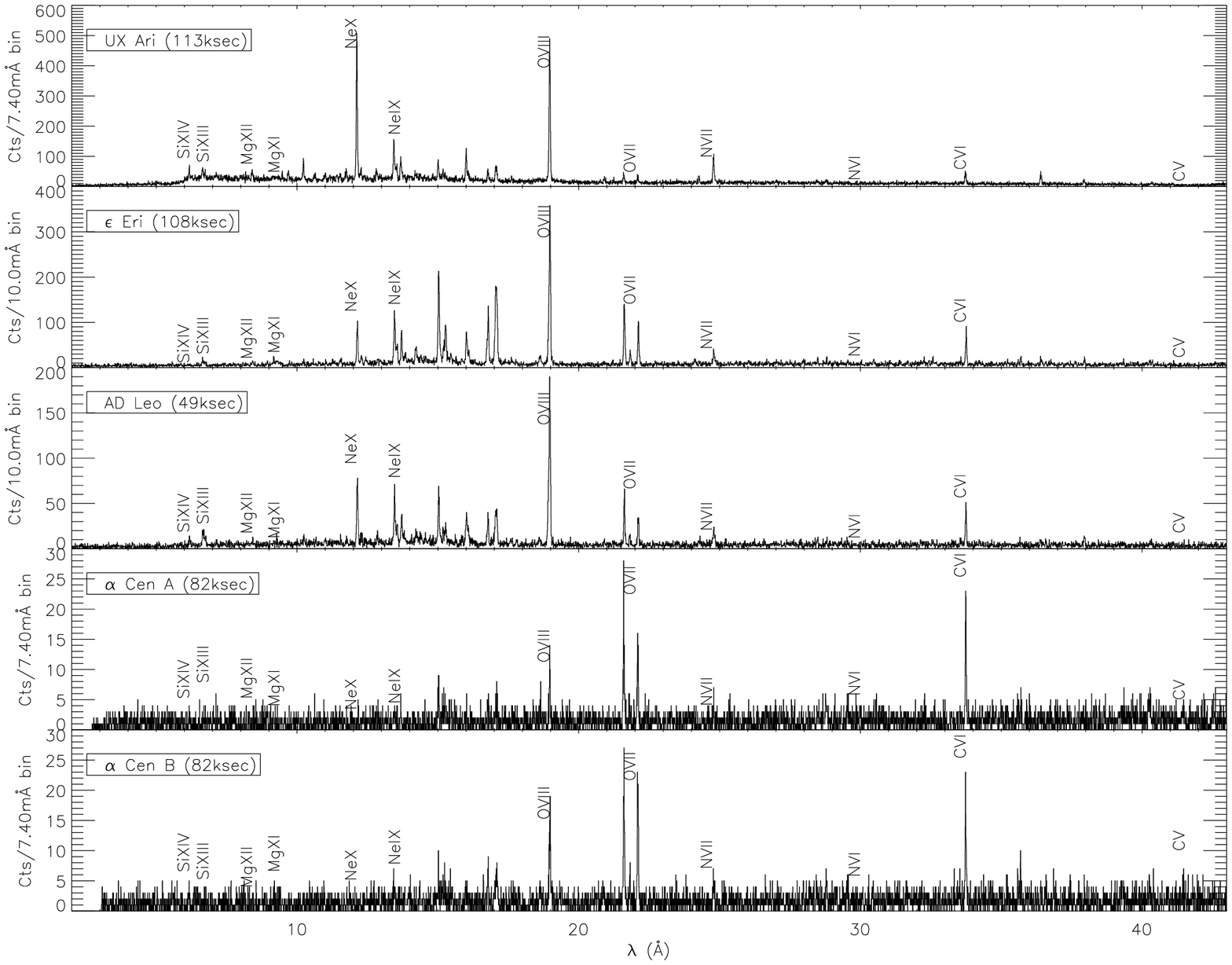}}}
\caption[]{\label{spec}The LETGS spectra of UX~Ari, $\epsilon$~Eri, AD~Leo, and
$\alpha$~Cen A and B in the range 1--40\,\AA\ with all lines used in this
paper marked. The instrumental background is subtracted.}
\end{figure*}

The new spectra obtained with the LETGS are shown in Fig.~\ref{spec} for the
stars UX~Ari, $\epsilon$~Eri, AD~Leo, and $\alpha$~Cen A and B in the
wavelength range 1--40\,\AA. The instrumental
background has been subtracted, i.e., the pure continuum plus line emission is
shown in units of counts per bin; no corrections for the effective areas of the
LETGS have been applied. As can be seen from an inspection of
Fig.~\ref{spec}, the LETGS spectra of our sample stars differ considerably.
Considerable continuum emission is detected for the active stars
UX~Ari, $\epsilon$~Eri, and AD~Leo in the wavelength range 10\,\AA\ - 20\,\AA,
where the effective LETGS area is largest, while for $\alpha$~Cen A and B
most of the emission is found in spectral lines. Strong emission lines from O,
Fe, Si, Ne, and Mg are seen for UX~Ari, AD~Leo, and $\epsilon$~Eri in the
wavelength range 10\,\AA\ -- 20\,\AA, while for $\alpha$~Cen A the O\,{\sc vii}
line at 21.6\,\AA\ is stronger than the O\,{\sc viii} line at 18.97\,\AA.
In the long wavelength region, the prominent Fe\,{\sc ix} at 171\,\AA\ is
found to be the strongest line in  $\alpha$~Cen A and B, while it
is not detected in the spectra of $\epsilon$~Eri, AD~Leo, and UX~Ari.
Instead we find highly ionized Fe ions from Fe\,{\sc xviii} (94\,\AA) up to
Fe\,{\sc xxiii} (e.g., 132\,\AA) in the UX~Ari spectrum but not for
$\epsilon$~Eri and AD~Leo.\\
From Fig.~\ref{spec} it is clear that isolated lines can be measured and line
ratios can be calculated.
In addition, broad band X-ray luminosities can be computed in the following
way: The number of photons in a wavelength bin is converted to an energy flux
by multiplying with $hc/\lambda$A$_{\rm eff}$ with $h$ the Planck constant,
$c$ the speed of light, A$_{\rm eff}(\lambda)$ the wavelength-dependent
effective area. Total X-ray luminosities in a given band pass are
obtained by summing all energy fluxes calculated within the corresponding bins
and dividing by $4\pi d^2$ ($d$ is distance). Surface
fluxes $F_X$ are calculated from $F_X=L_X/4\pi R_\star^2$ with $R_\star$
denoting the stellar radius. The X-ray luminosities and surface fluxes obtained
in this fashion are listed in Table~\ref{sprop} for all sample stars.

\subsection{Measurement of line fluxes}
\label{meas}

\begin{table*}
\caption[ ]{\label{flux_1}Measured line counts for the low temperature H-like
and He-like ions O, N, and C (cf. Sect.~\ref{meas})}
\renewcommand{\arraystretch}{1.2}
\begin{tabular}{r|cccccc}
&\multicolumn{2}{c}{O}&\multicolumn{2}{c}{N}&\multicolumn{2}{c}{C}\\
&\sc viii&\sc vii&\sc vii&\sc vi&\sc vi&\sc v\\
$\lambda$/\AA&18.97&21.6/21.8/22.1&24.74&28.8/29.1/29.5&33.74&40.3/40.7/41.5\\
A$_{\rm eff}$/cm$^2$&24.31&15.6/15.34/15.32&15.26&13.9/13.6/13.9&11.59&4.9/3.36/3.24\\
\hline
Algol (r)          & 2882.96 $\pm$ 57.81& 262.49 $\pm$ 22.6  &
1119.05 $\pm$ 38.38 & 141.33 $\pm$ 21.10 & $<$50 &--\\
(i)    & & 128.77 $\pm$ 18.70 & & 188.23 $\pm$ 25.53 & &-- \\
(f)    & & 120.90 $\pm$ 18.00 & & 37.50  $\pm$ 14.31 & &--\\
Capella (r)       & 14676.8 $\pm$ 124.3 & 3071.2 $\pm$ 56.0 & 2280.3 $\pm$ 53.6&
491.2 $\pm$ 31.49 & 2151.1 $\pm$ 51.3 & 440.7 $\pm$ 26.9\\
(i)    & & 544.8  $\pm$ 31.4  & & 228.2  $\pm$ 26.5 & & 101.3 $\pm$ 18.24\\
(f)    & & 2135.2 $\pm$ 51.1  & & 384.5  $\pm$ 29.4 & & 160.2 $\pm$ 22.37\\
Procyon (r)  & 673 $\pm$ 27.6 & 731.6 $\pm$ 28.7 & 206.9 $\pm$ 16.95&
200.2 $\pm$ 16.8 & 697.5 $\pm$ 28.1 & 203.8 $\pm$ 17.0\\
(i)    & & 203.0  $\pm$ 16.8  & & 77.4  $\pm$ 12.3  & & 123.1 $\pm$ 14.2 \\
(f)    & & 652.4  $\pm$ 27.3  & & 97.1  $\pm$ 13.2  & & 63.4  $\pm$ 13.5 \\
$\epsilon$~Eri (r) & 2025.8 $\pm$ 46.10 & 697.02 $\pm$ 27.70 &
178.85 $\pm$ 15.85 & 49.97  $\pm$  9.52 & 289.88 $\pm$ 18.66 & 51.83 $\pm$ 9.63\\
(i)    & & 153.61 $\pm$ 14.85 & & 22.86  $\pm$ 7.95 & & 16.42 $\pm$ 7.17\\
(f)    & & 453.99 $\pm$ 22.78 & & 24.41  $\pm$ 7.93 & & $<$12\\
$\alpha$~Cen~A (r) & 60.23  $\pm$  8.63 & 115.93 $\pm$ 11.46 &
17.67  $\pm$ 4.96  & 38.40  $\pm$  7.53 & 117.65 $\pm$ 11.44 & 39.38  $\pm$ 8.41\\
(i)    & & 23.57  $\pm$ 6.01  & & 16.79  $\pm$ 5.40 & & 6.4    $\pm$ 5.3\\
(f)    & & 89.68  $\pm$ 10.24 & & 34.63  $\pm$ 7.48 & & 12.98  $\pm$ 11.44\\
$\alpha$~Cen~B (r) & 124.18 $\pm$ 11.98 & 141.27 $\pm$ 12.55 &
32.07 $\pm$ 6.89   & 23.43  $\pm$  6.06 & 102.86 $\pm$ 11.26 & 26.71 $\pm$ 7.40\\
(i)    & & 24.42  $\pm$ 5.88  & & 10.39 $\pm$ 6.47  & & 6.47  $\pm$ 5.19 \\
(f)    & & 137.66 $\pm$ 12.34 & & 20.12 $\pm$ 6.10  & & 20.69 $\pm$ 7.51 \\
UX~Ari (r)        & 3207.85 $\pm$ 59.61 & 233.27 $\pm$ 21.11 & 602.85 $\pm$ 27.97& 96.88 $\pm$ 15.62 & 309.36 $\pm$ 20.95&--\\
(i)    & & $<$30              & & 27.17  $\pm$ 12.94 & & --\\
(f)    & & 156.50 $\pm$ 18.41 & & 36.13  $\pm$ 13.16 & & --\\
AD~Leo (r)& 1238.35 $\pm$ 36.40 & 263.97 $\pm$ 17.47 & 146.37  14.41 & 28.25 $\pm$ 7.89 & 203.41 $\pm$ 15.54 & 31.72 $\pm$ 8.44\\
(i)    & & 53.01 $\pm$  9.21 & & 19.85 $\pm$ 8.28 & & --\\
(f)    & & 170.31 $\pm$ 14.52 & & 19.91 $\pm$ 7.89 & & --\\
YY~Gem (r)& 971.47 $\pm$ 32.27&193.31 $\pm$ 15.47 & 115.51 $\pm$ 13.20 & -- & 149.44 $\pm$ 14.01 & --\\
(i)    & & 50.64 $\pm$ 9.63   & & -- & & --\\
(f)    & & 115.75 $\pm$ 12.44 & & -- & & --\\
HR\,1099 (r)& 5584.06 $\pm$ 78.60 & 470.36 $\pm$ 27.81 & 502.08 $\pm$ 29.3 & -- & 627.76 $\pm$ 29.63 & --\\
(i)    & & 114.74 $\pm$ 18.72 & & -- & & --\\
(f)    & & 254.09 $\pm$ 22.44 & & -- & & --\\
\hline
\end{tabular}
\begin{flushleft}
\renewcommand{\arraystretch}{1}
\end{flushleft}
\end{table*}

\begin{table*}
\caption[ ]{\label{flux_2}Measured line counts for the highly ionized H-like and
He-like ions Si, Mg, and Ne (cf. Sect.~\ref{meas})}
\renewcommand{\arraystretch}{1.2}
%{\scriptsize
\begin{tabular}{r|cccccc}
&\multicolumn{2}{c}{Si}&\multicolumn{2}{c}{Mg}&\multicolumn{2}{c}{Ne}\\
&\sc xiv&\sc xiii&\sc xii&\sc xi&\sc x&\sc ix\\
$\lambda$/\AA&6.18&6.65/6.69/6.74&8.42&9.17/9.23/9.31&12.14&13.45/13.55/13.7\\
A$_{\rm eff}$/cm$^2$&35.93&37.54/37.4/37.17&32.3&27.7/27.45/27.25&24.96&26.15/26.23/26.29\\
\hline
Algol (r)         & 658.32 $\pm$ 35.06 & 480.7 $\pm$ 35.95 & 578.04 $\pm$ 33.13&
 224.12 $\pm$ 26.15 & 2481.48 $\pm$ 56.31 & 631.23 $\pm$ 36.62\\
(i)    & & 86.40   $\pm$ 33.10 & & 84.85 $\pm$ 23.29 & & 80.52 $\pm$ 30.74\\
(f)    & & 314.10  $\pm$ 30.60 & & 76.09 $\pm$ 22.80 & & 347.91 $\pm$ 31.94\\
Capella (r)       & 679.72 $\pm$ 31.83 & 2487.72 $\pm$ 66.56 & 1682.28 $\pm$ 47.12 & 2654.49 $\pm$ 60.12 & 6479.41 $\pm$ 88.70 & 3960.85 $\pm$ 79.78\\
(i)    & & 307.16  $\pm$ 62.44 & & 481.71  $\pm$ 38.70 & & 1109.18 $\pm$ 64.76\\
(f)    & & 1346.19 $\pm$ 48.71 & & 1452.03 $\pm$ 47.24 & & 2312.77 $\pm$ 66.94\\
Procyon (r) & -- & -- & -- & -- & 10.67 $\pm$ 4.87 & 66.93 $\pm$ 10.46\\
(i)    & & -- & & -- & & 13.55 $\pm$ 6.63\\
(f)    & & -- & & -- & & 35.03 $\pm$ 8.47\\
$\epsilon$~Eri(r) & $<$19  & 62.81 $\pm$ 11.75 & 45.05 $\pm$ 9.96 &
 68.83 $\pm$ 11 & 480.78 $\pm$ 24.0 & 564.02 $\pm$ 27.20\\
(i)    & & 13.47 $\pm$ 10.6 & & 27.93  $\pm$ 9.1 & & 116.91 $\pm$ 17.64\\
(f)    & & 37.55 $\pm$ 9.82 & & 44.09  $\pm$ 9.8 & & 347.98 $\pm$ 22.35\\
UX~Ari (r)	  & 315.75 $\pm$ 26.17 & 321.13 $\pm$ 36.61 & 165.18 $\pm$ 20.70 & 47.54 $\pm$ 17.63 & 3038.04 $\pm$ 58.74 & 873.08 $\pm$ 35.98\\
(i)    & & 48.81  $\pm$ 39.98 & & $<$ 25            & & 187.82 $\pm$ 25.50\\
(f)    & & 146.42 $\pm$ 27.54 & & 29.62 $\pm$ 16.78 & & 465.67 $\pm$ 29.87\\
AD~Leo (r) & 52.06 $\pm$ 9.38 & 81.35 $\pm$ 12.93 & 17.18 $\pm$ 5.6 & -- & 407.87 $\pm$ 21.82 & 276.32 $\pm$ 19.20\\
(i)    & & 34.66 $\pm$ 12.31 & & -- & & 73.3 $\pm$ 13.69\\
(f)    & & 45.55 $\pm$ 10.02 & & -- & & 165.42 $\pm$ 16.04\\
YY~Gem (r) & 72.49 $\pm$ 11.06 & 82.50 $\pm$ 12.93 & 44.58 $\pm$ 10.0& 28.58 $\pm$ 8.7& 414.84 $\pm$ 22.26 & 240.53 $\pm$ 18.54\\
(i)    & & $<$ 10 & & $<$ 7 & & 30.67 $\pm$ 11.76\\
(f)    & & 31.01 $\pm$ 7.48 & & 9.4 $\pm$ 7.5 & & 137.45 $\pm$ 15.34\\
HR\,1099 (r) & 587.78 $\pm$ 33.17 & 458.55 $\pm$ 37.14 & 470.08 $\pm$ 35.0 & 372.84 $\pm$ 31.71 & 4421.27 $\pm$ 72.63 & 1231.66 $\pm$ 45.61\\
(i)    & & 90.74 $\pm$ 34.94 & & 110.08 $\pm$ 27.41 & & 191.75 $\pm$ 34.16\\
(f)    & & 262.68 $\pm$ 31.14 & & 188.34 $\pm$ 27.24 & & 727.60 $\pm$ 39.44\\
\hline
\end{tabular}
%}
\begin{flushleft}
\renewcommand{\arraystretch}{1}
\end{flushleft}
\end{table*}

\begin{table*}[!ht]
\caption[]{\label{contam}Modeling of contaminating Fe lines in the Ne\,{\sc ix}
triplet in comparison with the Fe\,{\sc xvii} line at 15\,\AA. The last column
contains the ratio of the hot 13.51\,\AA\ line with the cooler 15\,\AA\ line,
corrected for effective areas and converted to energy fluxes.}
\renewcommand{\arraystretch}{1.2}
\begin{tabular}{l|cccc|cr}
$\lambda$/\AA&13.51&13.65&13.79&13.83&15.01&(13.51)/(15.01)\\
A$_{\rm eff}$/cm$^2$&26.2&26.27&26.31&26.32&27.21&1.039\\
&Fe\,{\sc xix/xxi}&Fe\,{\sc xix}&Fe\,{\sc xix}&Fe\,{\sc xvii}&Fe\,{\sc xvii}&\\
\hline
Algol&585.68 $\pm$ 39.30&$<$29&134.69 $\pm$ 30.13&47.70 $\pm$ 28.39&1018.44 $\pm$ 38.93&0.66 $\pm$ 0,05\\
Capella&4046.6 $\pm$ 89.18&470.89 $\pm$ 49.98&2173.57 $\pm$ 73.68&1839.88 $\pm$ 70.43&23146.21 $\pm$ 161.52&0.20 $\pm$ 0.005\\
Procyon&&22.56 $\pm$ 7.52&&&105.94 $\pm$ 13.81&\\
$\epsilon$~Eri&147.26 $\pm$ 20.81&$<$13&34.10 $\pm$ 14.24&80.25 $\pm$ 15.6&1188.05 $\pm$ 36.92&0.14 $\pm$ 0.02\\
UX~Ari&195.87 $\pm$ 28.18&$<$20&51.58 $\pm$ 21.64&$<$30&383.02 $\pm$ 25.67&0.59 $\pm$ 0.09\\
AD~Leo&59.49 $\pm$ 14.84&$<$20&46.61 $\pm$ 11.65&$<$20&300.18 $\pm$ 19.52&0.23 $\pm$ 0.06\\
YY~Gem&91 $\pm$ 15.1&$<$16&22.4 $\pm$ 10.4&$<$20&234.29 $\pm$ 17.64&0.45 $\pm$ 0.08\\
HR\,1099&593.43 $\pm$ 42&$<$30&218 $\pm$ 33.3&77.9 $\pm$ 30.65&1059.2 $\pm$ 41&0.65 $\pm$ 0.05\\
\hline
\end{tabular}
\begin{flushleft}
\renewcommand{\arraystretch}{1}
\end{flushleft}
\end{table*}

The counts measured in emission lines are obtained using the CORA line
fitting tool, which has been developed and applied by Ness et al.
(\cite{ness_cap}); a detailed description with examples is given
by Ness \& Wichmann (\cite{new02}).
The raw counts obtained with CORA represent the fitted number of
expected counts for a given background. Measurement errors are given as
$1\,\sigma$ errors and include statistical errors
and correlated errors in cases of line blends, but do not include systematic
background errors. The raw counts obtained in this fashion
are listed in Table~\ref{flux_1} for the Ly$_{\alpha}$ lines and He-like
triplets of oxygen, nitrogen, and carbon, and in
Table~\ref{flux_2} for the Ly$_{\alpha}$ lines and He-like triplets of
neon, magnesium, and silicon. An inspection of
Table~\ref{flux_1} shows that the O\,{\sc viii} Ly$_{\alpha}$ line as well as
the resonance line in the O\,{\sc vii} triplet has been detected in all sample
stars. Nitrogen is detected in all stars, but N\,{\sc vi} has not been
detected in YY~Gem and HR\,1099. In Algol no carbon lines have been
detected (cf. Schmitt \& Ness \cite{schm02}), while the carbon Ly$_{\alpha}$
line at 33.74\,\AA\ has been detected in all other stars. The interpretation
of data on the C\,{\sc v} triplet is complicated; the effective area of
the LETGS is rather small in this region and further, for the active stars,
third order blending occurs with the neon triplet and Fe\,{\sc xvii/xix}
lines; still we find emission from the C\,{\sc v} resonance line at 40.3\,\AA\ 
in all cases but Algol, HR\,1099, and YY~Gem. In
Table~\ref{flux_2} we summarize our results for H-like and He-like lines of
neon, magnesium, and silicon. Note that in the (under-exposed) spectra of
$\alpha$~Cen A and B none of these lines has been detected, while in the much
better exposed spectrum of Procyon the Ne\,{\sc ix} triplet is clearly seen,
but Ne\,{\sc x} is very weak. In all other stars we find emission from both the
H-like and He-like ions from neon and silicon, however, magnesium is very
weak in most of the stars.\\

\subsubsection{Si\,{\sc xiii} and Mg\,{\sc xi} triplets}

\begin{figure}[!ht]
  \resizebox{\hsize}{!}{\rotatebox{90}{\includegraphics{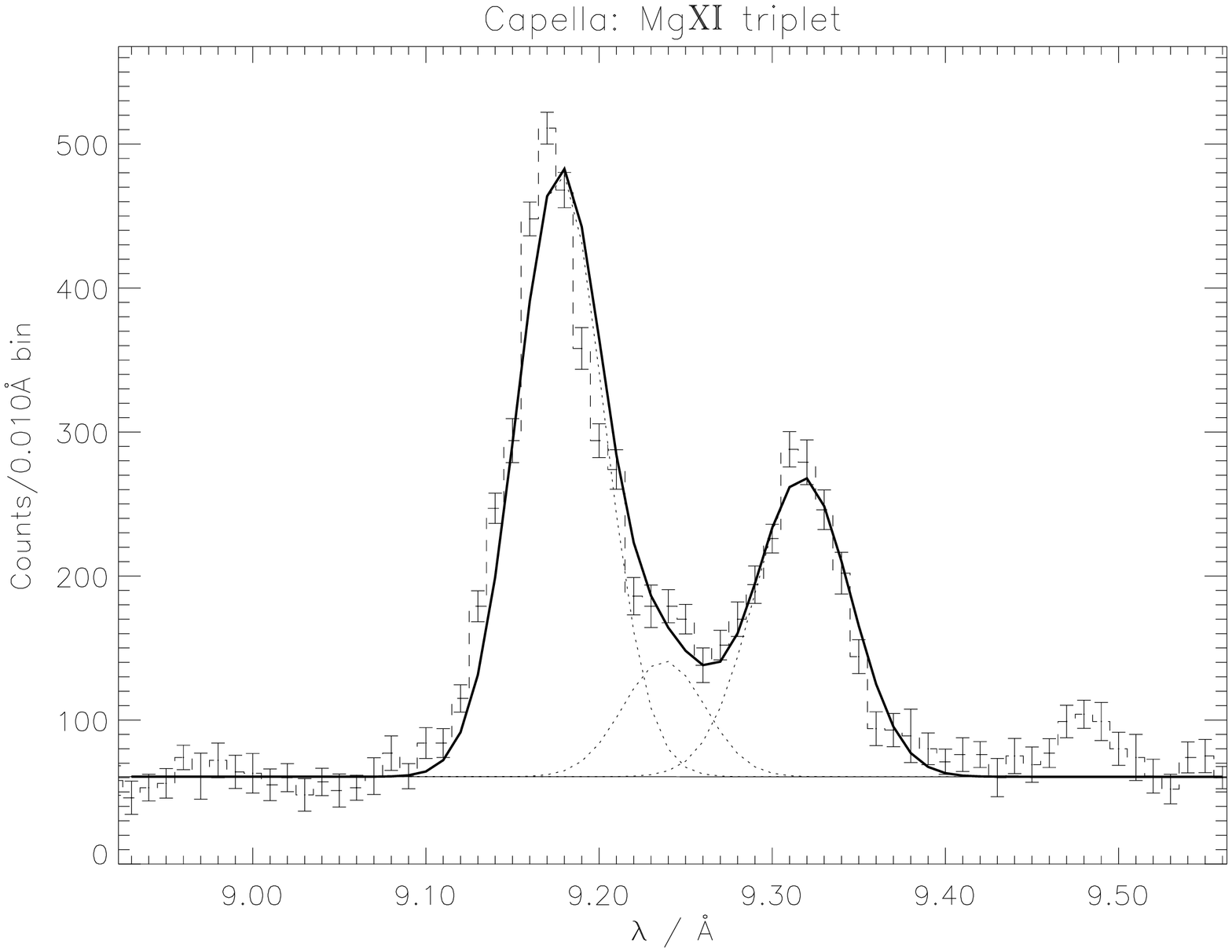}}}
\caption[]{\label{mg}Measured spectrum of the Mg\,{\sc xi} triplet for Capella
with the best fit.}
\end{figure}

The Si\,{\sc xiii} and the Mg\,{\sc xi} triplets at 6.7\,\AA\ and 9.2\,\AA\ are
formed at temperatures of 7\,MK and 6.8\,MK and are density sensitive at
densities log$(n_e)>10^{12}$\,cm$^{-3}$. In contrast to the HETGS with its
larger effective areas and higher resolution, these triplets are not fully
resolved in the LETGS spectra, but the blend can be modeled by a set of three
partially overlapping lines with fixed wavelength differences. For example,
the Si\,{\sc xiii} triplet was modeled by Ness et al. (\cite{ness_alg}) and
a plot is shown there. As can be seen from Table~\ref{flux_2} the
Mg\,{\sc xi} triplet is not detected
for AD~Leo and is very weak in UX~Ari and YY~Gem. Fig.~\ref{mg} demonstrates
this procedure for the
Mg\,{\sc xi} triplet for Capella with our best fit. From this plot it can
be seen that the line blend can be well modeled with the three lines for
Mg\,{\sc xi}. All our results quoted in this paper on the Si\,{\sc xiii} and
the Mg\,{\sc xi} triplets have been obtained with such a de-blending procedure.

\subsubsection{Ne\,{\sc ix} triplet}

\begin{figure}[!ht]
  \resizebox{\hsize}{!}{\rotatebox{90}{\includegraphics{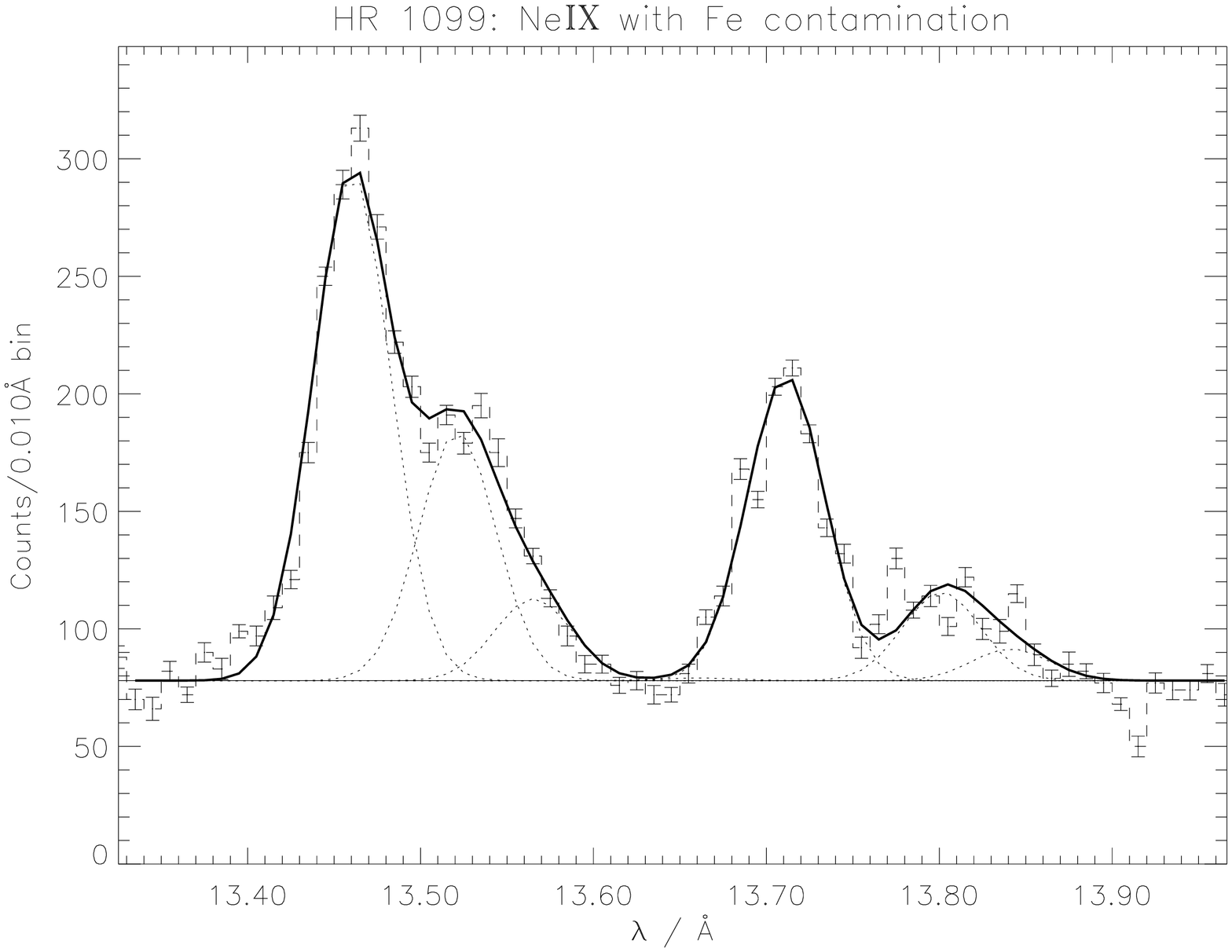}}}
  \resizebox{\hsize}{!}{\rotatebox{90}{\includegraphics{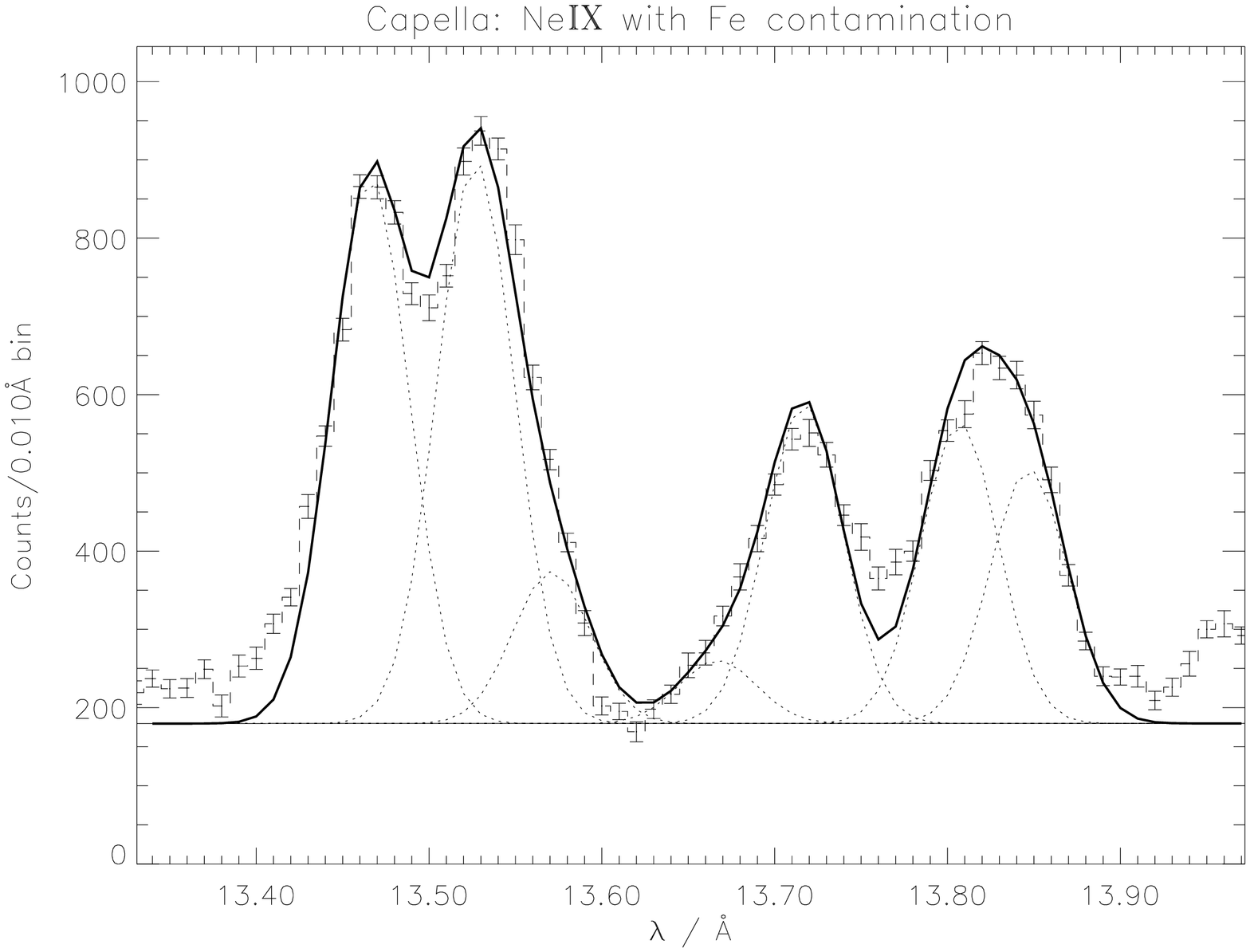}}}
\caption[]{\label{nehr}Measured spectrum of the Ne\,{\sc ix} triplet with
Fe contamination for HR\,1099 (top) and Capella (bottom) with the best fits
using seven wavelength values (cf. Table~\ref{contam}) iterated only as a whole
multiplet complex. The Gaussian line width is fixed to $\sigma=0.0222$\,\AA\ for all lines.}
\end{figure}

\begin{figure}[!ht]
 \resizebox{\hsize}{!}{\includegraphics{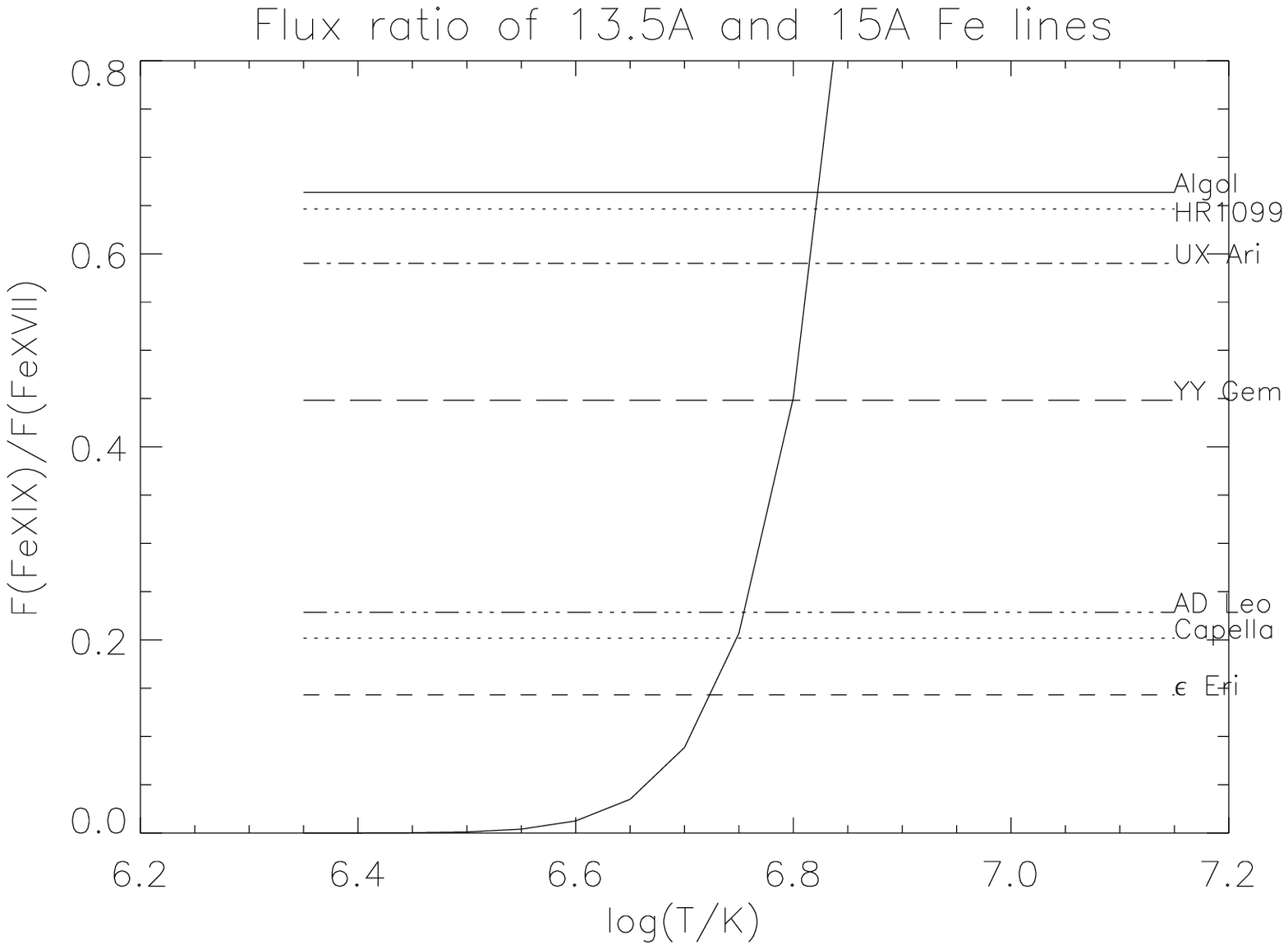}}
\caption[]{\label{cont_fig}Comparison of flux ratios derived from the
13.51\,\AA\ line (which contaminates the Ne\,{\sc ix} intercombination line)
and the strong Fe\,{\sc xvii} line at 15\,\AA\ with the MEKAL ratio (solid
curve). The theoretical
13.52\,\AA\ flux is derived assuming the MEKAL lines at 13.504\,\AA\ and
13.52\,\AA\ to contribute with their individual fluxes.}
\end{figure}

The analysis of the Ne\,{\sc ix} triplet is complicated because of severe
blending of the intercombination line with Fe\,{\sc xix} at 13.51\,\AA. An
approach of still obtaining useful results is introduced for Algol by Ness et
al. (\cite{ness_alg}), where the G-ratio (f+i)/r is fixed to 0.8 as is
expected for a collision dominated plasma with a reasonable temperature. In
this paper we follow a different, more empirical approach. In an HEG spectrum
of Capella (Ness et al. \cite{nessb02}) a total of 18 lines can be
identified, which we combine - for the purposes of LETGS modeling - into a set
of four contaminating lines (which are listed in Table~\ref{contam}). These
contamination lines are due to Fe\,{\sc xvii} and Fe\,{\sc xix}.
With these seven lines, i.e., four lines from Fe\,{\sc xvii/xix} and the
three Ne\,{\sc ix} He-like triplet lines we calculate a best fit. The relative
line positions as well as the line widths ($\sigma=0.0222$\,\AA)
are kept constant during the iteration. The fit results for
HR\,1099 and Capella are shown in Fig.~\ref{nehr}. Obviously a good
fit can be obtained with the wavelengths estimated from the Capella HEG
spectrum for both stars. The two double peaks visible in the Capella spectrum
appear in a different form in the HR\,1099 spectrum. We attribute this
difference to an anomalously high Ne abundance in HR\,1099 (cf., e.g., Brinkman
et al. \cite{brink01}). In the spectrum of HR\,1099 we additionally
identified Ne\,{\sc x} Ly$_{\alpha, \beta, \gamma,\ {\rm and}\ \delta}$ with
45.35 $\pm$ 0.74, 6.04 $\pm$ 0.34, 2.33 $\pm$ 0.27, and 1.96 $\pm$ 0.26
cts/ksec and for Capella with 20.65 $\pm$ 0.33, 2.69 $\pm$ 0.15, 1.04 $\pm$
0.12, and 0.88 $\pm$ 0.12 cts/ksec, respectively. The count rates differ by a
factor of more than two in favor of HR\,1099 suggesting higher Ne abundances
in HR\,1099 compared to Capella. We consider temperature effects less
important, at least not accounting for this large discrepancy.\\
As a cross check we compare the strengths of the contaminating line at
13.51\,\AA\ with the
strong isolated Fe\,{\sc xvii} line at 15\,\AA. In Table~\ref{contam} we show
the fit results for all contaminating lines and for the 15\,\AA\ line. In the
last column we calculated the flux ratios of the 13.51\,\AA\ line and the
15\,\AA\ line. These ratios are plotted in Fig.~\ref{cont_fig} in comparison
with the temperature dependent theoretical curve as calculated with MEKAL
(MEwe, KAastra, \& Liedahl \cite{mewe95})
assuming the 13.51\,\AA\ line to consist of two Fe\,{\sc xix} lines at
13.504\,\AA\ and 13.52\,\AA. Under this assumption the ratio (13.51)/(15.01)
must be temperature sensitive since Fe\,{\sc xix} is formed at higher
temperatures than Fe\,{\sc xvii}. From Fig.~\ref{cont_fig} we see a trend
suggesting Algol, HR\,1099, and UX~Ari to contain the hottest plasma in our
sample. This is consistent with our temperature measurements in
Tables~\ref{restab1} to \ref{restab3} as well as with the trend seen in
Fig.~\ref{LyLx}. The ratios for the other stars seem also very reasonable.
We finally point out that the G-ratios derived from our Ne measurements are all
consistent with the expected value of 0.8, which again demonstrates internal
self-consistency of our procedure. \\
Another problem with Ne is the blending of H-like Ne\,{\sc x} at 12.14\,\AA\
with Fe\,{\sc xvi}/Fe\,{\sc xvii} ($\lambda$=12.134\,\AA). The contamination
with the Fe lines is most critical at temperatures less than 4\,MK, and can be
neglected at temperatures higher than 6\,MK.\\

\subsubsection{O\,{\sc vii} triplet}

\begin{figure}[!ht]
  \resizebox{\hsize}{!}{\rotatebox{270}{\includegraphics{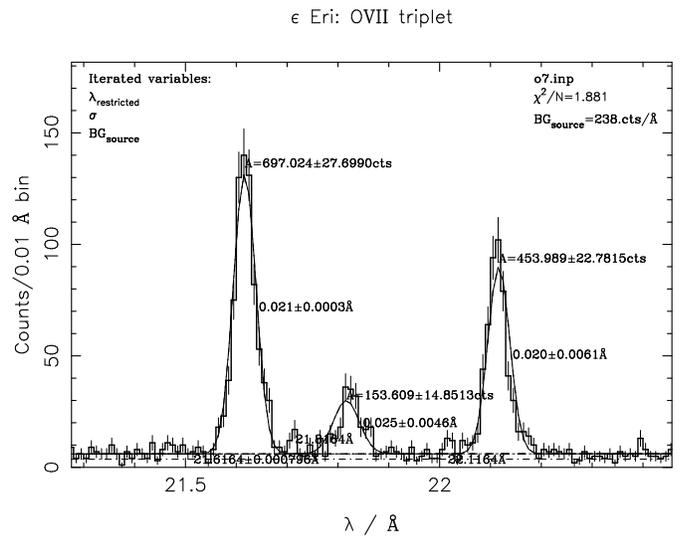}}}
\caption[]{\label{o7eps}Measurement of the O\,{\sc vii} triplet for
$\epsilon$~Eri (105~ksec).}
\end{figure}

The most prominent He-like triplet is the O\,{\sc vii} triplet which is
detected in all stars. In Fig.~\ref{o7eps} we show the region around the
O\,{\sc vii} triplet for $\epsilon$~Eri; the resonance line, the
intercombination line, and the forbidden line are all clearly detected above
the weak continuum. A weak line at 21.7\,\AA\ may be present, but it does not
compromise the measurement of the O\,{\sc vii} line fluxes. For
the f/i-ratio for $\epsilon$~Eri we find a value of 2.96 $\pm$ 0.32, which is
not consistent with the low density limit of 3.95.

\begin{figure}[!ht]
  \resizebox{\hsize}{!}{\rotatebox{90}{\includegraphics{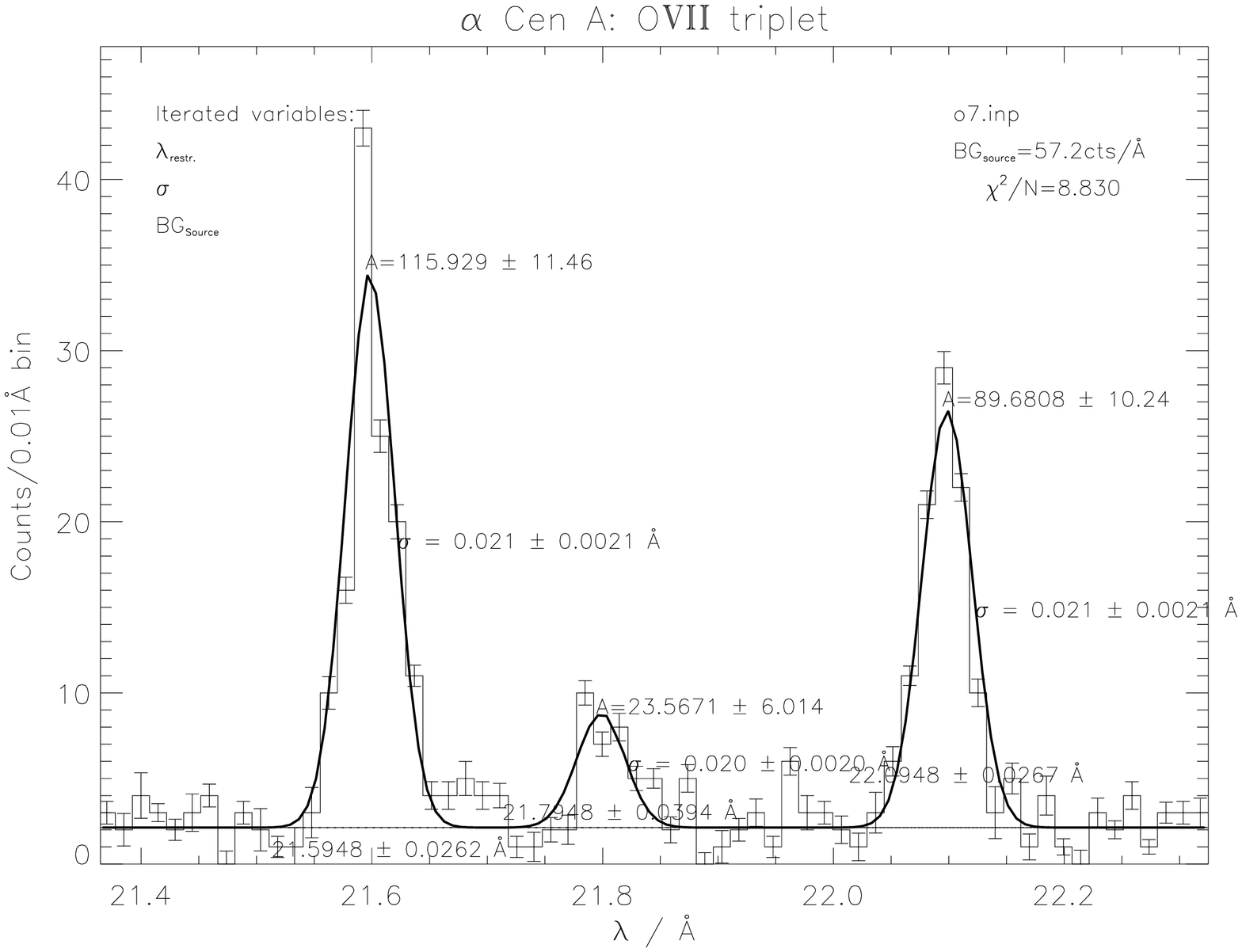}}}
  \resizebox{\hsize}{!}{\rotatebox{90}{\includegraphics{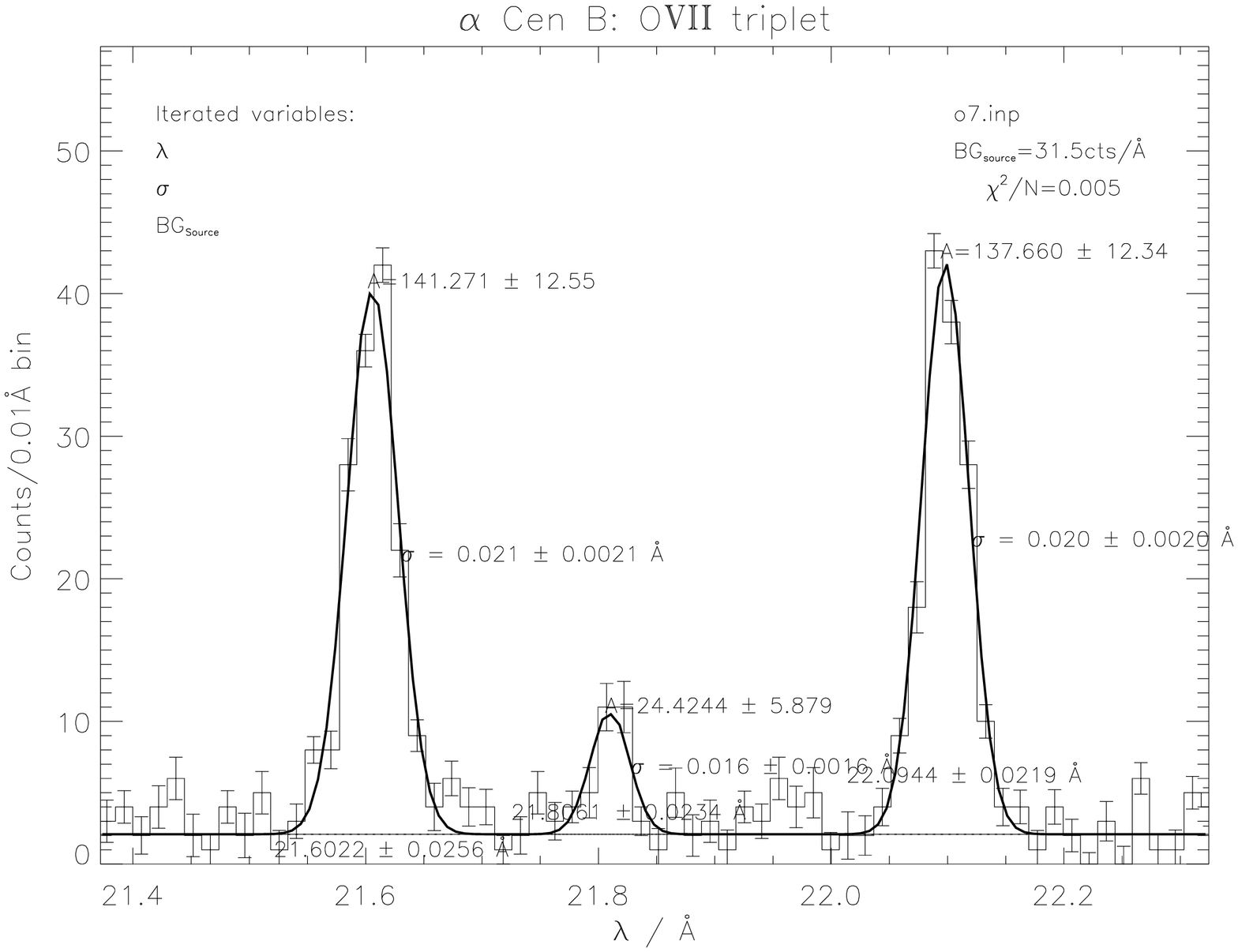}}}
\caption[]{\label{o7acen}Measurement of the O\,{\sc vii} triplet for
$\alpha$~Cen~A (top) and B (bottom) (81.5~ksec).}
\end{figure}

In Fig.~\ref{o7acen} we plot the O\,{\sc vii} triplets for
$\alpha$~Cen A and B, which can easily be spectrally resolved because
of the superb angular resolution of the {\it Chandra} mirrors. These spectra
have lower SNR than the $\epsilon$~Eri spectrum and we measure f/i = 3.81
$\pm$ 1.06 and f/i = 5.6 $\pm$ 2.9, i.e., numbers which are fully consistent
with the low density limit.

\begin{figure}[!ht]
  \resizebox{\hsize}{!}{\rotatebox{90}{\includegraphics{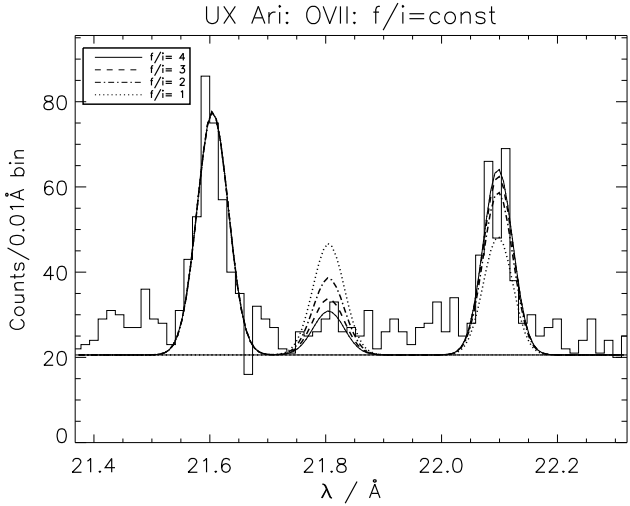}}}
\caption[]{\label{o7ux}Measured spectrum of the O\,{\sc vii} triplet for UX~Ari
with best fits constrained to satisfy f/i=4, 3, 2, and 1.}
\end{figure}

In Fig.~\ref{o7ux} we show the O\,{\sc vii} triplet for UX~Ari. The
corona of UX~Ari is much hotter than that of $\epsilon$~Eri and
$\alpha$~Cen A and B. Therefore the continuum is much higher so that the
weakest line of the O\,{\sc vii} triplet, the intercombination line, is not
detected, while the resonance line and the forbidden line are. This
implies that UX~Ari shows different properties compared to Algol
where intercombination and forbidden line have approximately the same
strength (cf. Ness et al. \cite{ness_alg}). To better illustrate the
observational situation
we plot in Fig.~\ref{o7ux} the recorded UX~Ari LETGS spectrum with
f/i-ratios of 4, 3, 2, and 1 superimposed. Clearly, the LETGS
measurements are consistent with the low density limit, but f/i-ratios as low
as f/i $\,\sim\,$2 cannot be excluded.\\

\subsubsection{N\,{\sc vi} triplet}

The N\,{\sc vi} triplet at 28.8, 29.1, and 29.5\,\AA\ has been detected
in all stars but YY~Gem (cf. Stelzer et al. \cite{stelz02}) and HR\,1099.
In all cases do we find f/i-values considerably below the low density
limit $R_{0,N} = 6.0$, the largest measurement being 2.19 $\pm$ 0.86 for
$\alpha$~Cen A and the smallest one being 0.21 $\pm$ 0.09 for Algol. The
interpretation of the N\,{\sc vi} triplet must take possible radiation
effects into consideration (cf. Sect.~\ref{radfield}). However, for cool stars
such as AD~Leo, YY~Gem, and $\epsilon$~Eri these radiation effects are
very small and the low f/i-ratio can unambiguously be attributed to
a density effect.

\subsubsection{C\,{\sc v} triplet}

The analysis of the C\,{\sc v} triplet at 41\,\AA\ is intimately connected with
measuring the Ne\,{\sc ix} triplet and the Fe contamination, which appears
superimposed on C\,{\sc v} in
third order. Fortunately, for the cool coronae of $\alpha$~Cen A and B and
Procyon emission from Ne\,{\sc ix} and Fe\,{\sc xvii} ions is either weak or
absent in first order (and particularly in third order which is damped by a
factor $\approx$ 20), so that C\,{\sc v} can be directly measured. For those
stars emission from the C\,{\sc v} triplet is detected; the data for Procyon
are discussed by Ness et al. (\cite{ness_cap}), for both components of
$\alpha$~Cen the SNR is so low that the intercombination line is only
marginally detected at best, while the resonance and forbidden lines are. Thus
detailed density diagnostics are hardly called for. For all other stars the
forbidden line of C\,{\sc v} is severely contaminated by third
order lines at 13.8\,\AA. This is a specific
disadvantage of the LETGS, however, no other instrument can
measure the C\,{\sc v} triplet. Since the first order signal is also
recorded by the same instrument, it is still possible
to model the C\,{\sc v} triplet taking the third
order contamination explicitly into account. This was demonstrated
by Ness et al. (\cite{ness_cap}) for Capella, however,
such a procedure requires high SNR data, which are unfortunately
available only for Capella. Since in addition in the active stars
Algol, UX~Ari, and HR\,1099 the carbon lines are absent or weak (cf. Schmitt
\& Ness \cite{schm02}), we therefore refrained from carrying out
a similar modeling procedure for the other sample stars.

\section{Results}
\label{result}

In Tables~\ref{restab1} to \ref{restab3} we provide a summary of the
plasma diagnostic results of our analysis
for the He-like ions Si\,{\sc xiii}, Mg\,{\sc xi}, Ne\,{\sc ix}, O\,{\sc vii},
N\,{\sc vi}, and C\,{\sc v}. Some ions are not detected as, e.g., C\,{\sc v}
is not detected in the hot stars Algol and UX~Ari, or Si\,{\sc xiii} is not
detected in the cool coronae of Procyon and $\alpha$~Cen A and B. The
ratio of H-like to He-like, Ly$_\alpha$/r, can not always be measured, since
the He-like triplet cannot always be fully resolved for silicon and magnesium.
For those cases the ratio Ly$_\alpha$/(r+i+f)=Ly$_\alpha$/He is also given. In
the following we describe how temperatures and densities are derived.

\subsection{Influence of radiation fields}
\label{radfield}

\begin{figure}[!ht]
  \resizebox{\hsize}{!}{\includegraphics{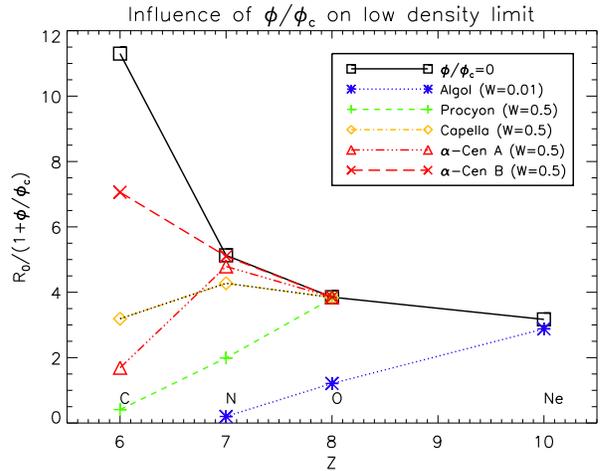}}
\caption[]{\label{radplot}Influence of different radiation fields on the
low-density limit restricting the density diagnostics. $W$ is the dilution
factor applied accounting for
the geometry of the stars. Plotted is the expected f/i ratio for the case
$n_e=0$ (cf. Eq.~1) versus atomic number Z.}
\end{figure}

Our sample contains stars of type B to M, which differ dramatically in
their photospheric radiation fields. Specific attention must therefore
be paid to the presence of such fields which
are different for the different He-like lines as well as in our sample
stars. For the stars with hotter photospheric
temperatures, the effects of radiation fields
on the low-Z ions cannot be neglected (e.g., Ness et al. \cite{ness_StC_p},
\cite{ness_CS}). In our sample the stars Procyon, Capella, and Algol are of
particular interest in that respect. For Procyon and Capella
we use the results obtained by Ness et al. (\cite{ness_cap}) and for Algol
the results obtained by Ness et al. (\cite{ness_alg}). In the case of Algol we
note that with the effects of the radiation field, which originates from the
companion B star, we assume a worst case scenario with the X-ray corona
fully immersed in the B-star's radiation field; Ness et al. (\cite{ness_StC_p})
discuss the possible geometrical configurations of the binary at the time of
the observation. For the K stars (cf. Table~\ref{sprop}) the effects of radiation
fields can be neglected. For $\alpha$~Cen A and B we measured the radiation
fields with IUE, corresponding to Ness et al. (\cite{ness_cap}), and found that
the radiation affects only the C\,{\sc v} triplet (cf. Table~\ref{restab1}
and Ness et al. \cite{ness_CS}). All the results of the radiation
fields measured with IUE are illustrated in Fig.~\ref{radplot}, where the
influence is given in terms of corrected low density limits.\\

\subsection{Densities}

The results of our measurements of the ratio of forbidden and intercombination
lines in the He-like triplets of Si, Mg, Ne, O, N, and C are listed in
Tables~\ref{restab1} to \ref{restab3}.
In our analysis we attribute all measured counts to the
respective intercombintion and forbidden lines; possible contamination by
other coincidental lines or dielectronic satellite lines will be discussed
in Sect.~\ref{conclusio}. For the purpose of discussion we assume
the low-density case and ask to what extent and at what confidence we can
exclude the low-density case on the basis of our LETGS measurements.
Specifically, we consider only those cases where the low-density limit can be
excluded at least on the ''2\,$\sigma$'' level. For Si the
only measurement indicating deviations from the low-density limit are those for
AD~Leo. The Si triplet is clearly broadened, yet the overall signal is rather
low. We also modeled the Si\,{\sc xiii} triplet with the constraint of f/i=2.6,
i.e., the low density limit, and found this fit only slightly worse. We
therefore conclude that any claim for a deviation of the Si\,{\sc xiii}
triplet from the low-density limit needs confirmation from an MEG or HEG
measurement. For the Mg triplet no deviation from the low-density limit above
the 2\,$\sigma$ level can be observed. For neon the only two
stars with significant deviations from the low density limit are UX~Ari and
Capella. In the latter case our measurement, i.e., 2.08 $\pm$ 0.14,
deviates significantly from the low density limit of 3.5. This is
puzzling since the corresponding measurement for the O\,{\sc vii} triplet is
consistent with the low density limit. From the higher resolution {\it Chandra}
MEG measurements of Ne\,{\sc ix} Ayres et al. (\cite{ayres01}) find f/i=2.9
$\pm$ 0.5 for Ne\,{\sc ix} and 2.6 $\pm$ 0.3 for O\,{\sc vii}. From an
apparently different {\it Chandra} MEG spectrum of Capella Phillips et al.
(\cite{phil01}) find an f/i-ratio of 2.64 $\pm$ 0.56 for O\,{\sc vii} in
agreement with the values reported by Ayres et al. (\cite{ayres01}), while for
Ne\,{\sc ix} these authors argue that the agreement between the {\it Chandra}
data and a low-density synthetic model spectrum is reasonable. Neither number is
consistent with our measurements, we note, however, that the MEG measurements
of Si, Mg, Ne, and O in HR\,1099 analyzed by Ayres et al. (\cite{ayres01}) are
all consistent with our results.
Also, Ness et al. (\cite{nessb02}) find a Ne\,{\sc ix} f/i ratio for
Capella consistent with our results using all HEG measurements available for
Capella. While our O\,{\sc vii}-ratio for Capella
is consistent with the XMM-RGS measurement for Capella reported by Audard et al.
(\cite{aud2}), a lower f/i ratio is found only with the MEG measurements.
This discrepancy is only found for Capella but not for HR\,1099. Concerning the
discrepancy with Ne\,{\sc ix} for Capella we point out that the blending
problems are more severe with the LETGS than with the MEG, such that the
higher f/i-ratios for Ne\,{\sc ix} measured with the MEG seem to be more
reliable.
However, the HEG has a better resolution than the MEG, such that
the spectra analysed by Ness et al. (\cite{nessb02}) should be most reliable.
Obviously, the best data are available for the oxygen triplet; for Algol,
Procyon, YY~Gem, $\epsilon$~Eri, and HR\,1099 do we find significant deviations
from the low-density limit and interpret those results in Sect.~\ref{conclusio}.\\
For the nitrogen triplet we find deviations from the low density limit
for all stars where sensitive measurements could be made; note that for HR\,1099
and YY~Gem no clear detections of the N triplet are available. Our discussion
in Sect.~\ref{radfield} indicates that for Algol and Procyon the low f/i-ratios
can be attributed to the radiation field alone, in all other sample stars we
must attribute the low f/i-ratios to high densities. As to the C\,{\sc v}
triplet, most of the LETGS data has very low SNR. The data for Procyon and
$\alpha$~Cen A and B cannot be reconciled with the low-density limit and
neither those for Capella (cf. Ness et al. \cite{ness_cap}), where the higher
order contamination can be well modelled. In all other cases we can make no
particularly sensitive statements except noting that in AD~Leo and possibly
YY~Gem the C\,{\sc v} resonance line is detected and far higher SNR is
required for any definite conclusions.

\subsection{X-ray luminosities and Temperatures}

\begin{figure*}[!ht]
  \resizebox{\hsize}{!}{\includegraphics{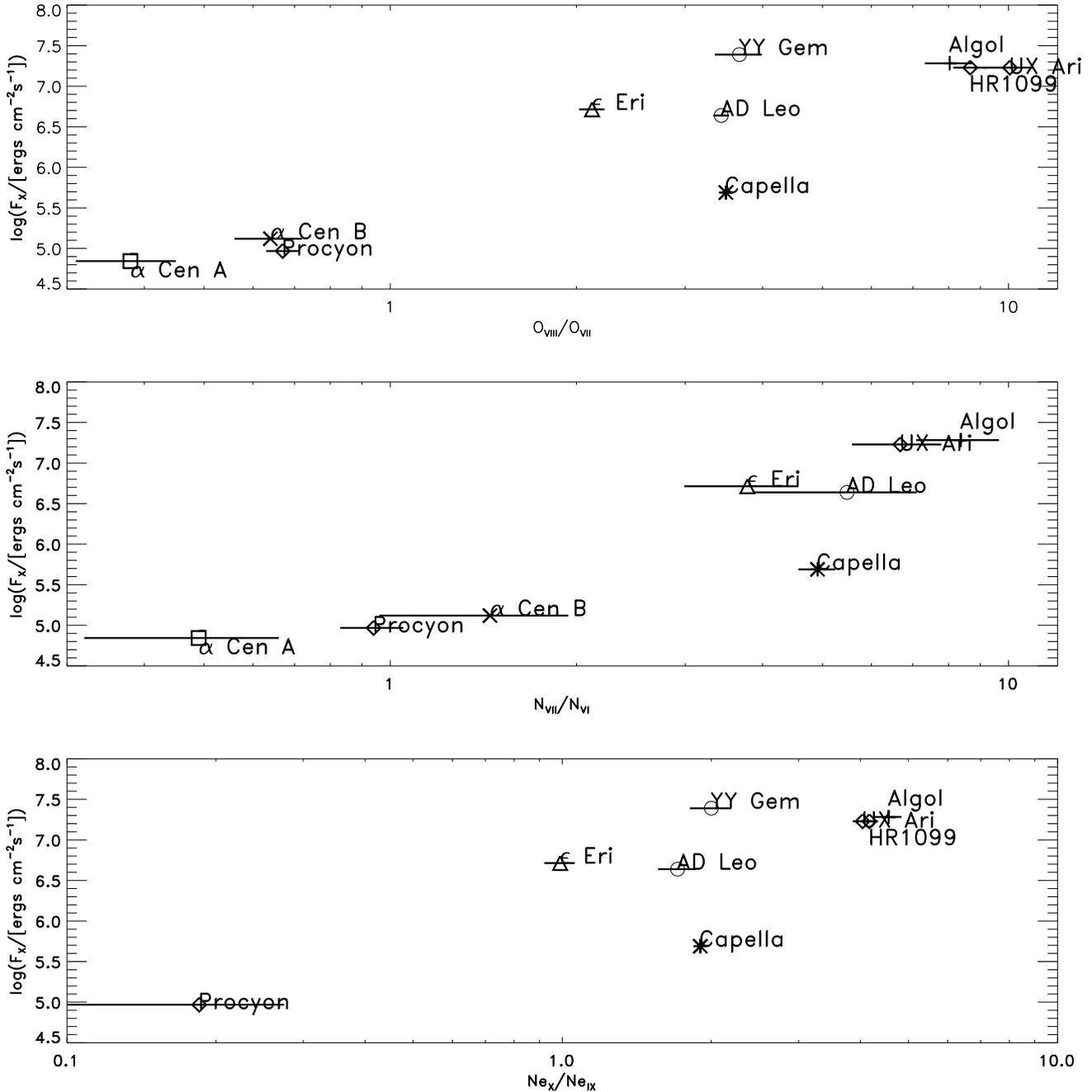}}
 \caption{\label{LyLx}
Relation between line ratio Ly$_\alpha$/He-like and surface flux
$F_X=L_X/4\pi R_\star^2$, with the stellar radii $R_\star$ in cgs units, for
O\,{\sc viii}/O\,{\sc vii} (top panel), N\,{\sc vii}/N\,{\sc vi}
(middle panel), and Ne\,{\sc x}/Ne\,{\sc ix} (bottom panel).}
\end{figure*}

As already discussed in Sect.~\ref{spectra} and listed in Table~\ref{sprop} we
directly measured X-ray luminosities by evaluating the total energy
flux from all photons recorded in a given wavelength bin and summing up.
We note in passing that this procedure underestimates the true energy flux
since photons from higher dispersion orders are treated incorrectly, however,
the overall error introduced by this is rather small. In Tables~\ref{flux_1}
and \ref{flux_2} we list the Ly$_{\alpha}$-lines and the resonance lines of
the He-like ions. Oxygen Ly$_{\alpha}$ and the oxygen He-like resonance lines
are detected for all our sample stars, while the corresponding lines for
nitrogen and neon are detected for almost all stars in our
sample. The measured X-ray luminosities $L_X$ range from $\approx 10^{27}$
up to $10^{32}$\,erg/sec while the measured
line ratios between the Ly$_{\alpha}$ and the He-like resonance line
range from values below unity for the inactive stars Procyon and
$\alpha$~Cen A and B up to 10 for the most luminous stars Algol, UX~Ari, and
HR\,1099. A similar trend is found when the ratios of the nitrogen
Ly$_{\alpha}$ line and the He-like resonance line as well as the corresponding
Ne ratios are considered. In Fig.~\ref{LyLx} we plot the mean X-ray surface flux
$F_X$, calculated as $L_X/4 \pi R_\star^2$ with the stellar radii $R_\star$
taken from Table~\ref{sprop}, for the line ratios O\,{\sc viii}/O\,{\sc vii},
N\,{\sc vii}/N\,{\sc vi}, and Ne\,{\sc x}/Ne\,{\sc ix}. We refrain from
carrying out a formal correlation analysis, but point out that larger
mean X-ray surface fluxes tend to go together with larger line ratios between
the Ly$_{\alpha}$ lines and the He-like resonance lines (cf. Fig.~\ref{LyLx}).
For each atomic species the ratio between the Ly$_{\alpha}$ and He-like
resonance
line is a monotonic function of temperature. Larger line ratios therefore imply
larger temperatures in the sense that the temperatures are weighted with the
corresponding emission measures. Different atomic species are sensitive to
emission measure in different temperature ranges and therefore represent different
``effective'' temperatures. Therefore Fig.~\ref{LyLx} demonstrates a
temperature-activity relationship. Such relationships between plasma
temperature and X-ray luminosity are not entirely new (e.g., Schrijver et al.
\cite{schr84}, Schmitt \cite{schm90}), but measuring these
line ratios as an indicator
of temperature is new. The measured line ratios can also be directly
converted into temperature if an isothermal plasma is assumed. We computed
temperatures by, first, comparing the measured line ratio of Ly$_\alpha$ lines
with the recombination line of the corresponding He-like ions with a temperature
dependent theoretical ratio taken from Mewe et al. (\cite{mewe85});
alternatively we can use the total triplet
emission for those cases in which the triplet is not fully resolved.
An alternative way consists of using the He-like G ratio, G=(f+i)/r.
In our analysis we compare the measured G ratios with the tables from
Porquet et al. (\cite{por01}), and interpolate quadratically. We always assume
low densities, and point out that the temperature is not very sensitive to
different densities. The temperatures derived in this fashion are quoted
in Tables~\ref{restab1} to \ref{restab3}. Inspection of Tables~\ref{restab1} to
\ref{restab3} shows first that the temperatures derived with the two methods do
not agree with the temperatures derived from the G-ratio in general being
smaller than those derived from the Ly$_{\alpha}$ lines, and
second, that different ions result in different temperatures, thus demonstrating
that more than one temperature component must be present. We emphasize in
this context that the above conclusions are independent of the elemental
abundances since we use only ratios of the same species.
\\
We also point out that the different stars ``cluster'' in certain regions of
the $F_X$ vs. Ly$_{\alpha}$/r-ratio plane. This ``clustering'' is best apparent
for oxygen where most measurements are available. The inactive stars $\alpha$
Cen A and B and Procyon form one group with log $F_X$ $\approx$ 5 and O\,{\sc
viii}/O\,{\sc vii} $\approx$ 0.5. The active late-type dwarfs AD~Leo, YY~Gem,
and $\epsilon$~Eri form another group with log $F_X$ $\approx$ 7 and O\,{\sc
viii}/O\,{\sc vii} $\approx$ 3, the very active giants HR\,1099, UX~Ari, and
Algol a third one with log $F_X$ $\approx$ 7.5 and O\,{\sc viii}/O\,{\sc vii}
$\approx$ 9; the RS\,CVn-like binary Capella appears low in these diagrams
because of the large radii (cf. Table~\ref{sprop}), which lead to low mean
X-ray surface fluxes. Whether or not this grouping
represents a physical effect or is merely due to chance is difficult to tell
in view of the sample of stars available. More observations covering a
different variety of stars are clearly required.
\\
\subsection{Correlation between density and surface flux}

\begin{figure*}[!ht]
\vspace{-2cm}
  \resizebox{\hsize}{!}{\rotatebox{90}{\includegraphics{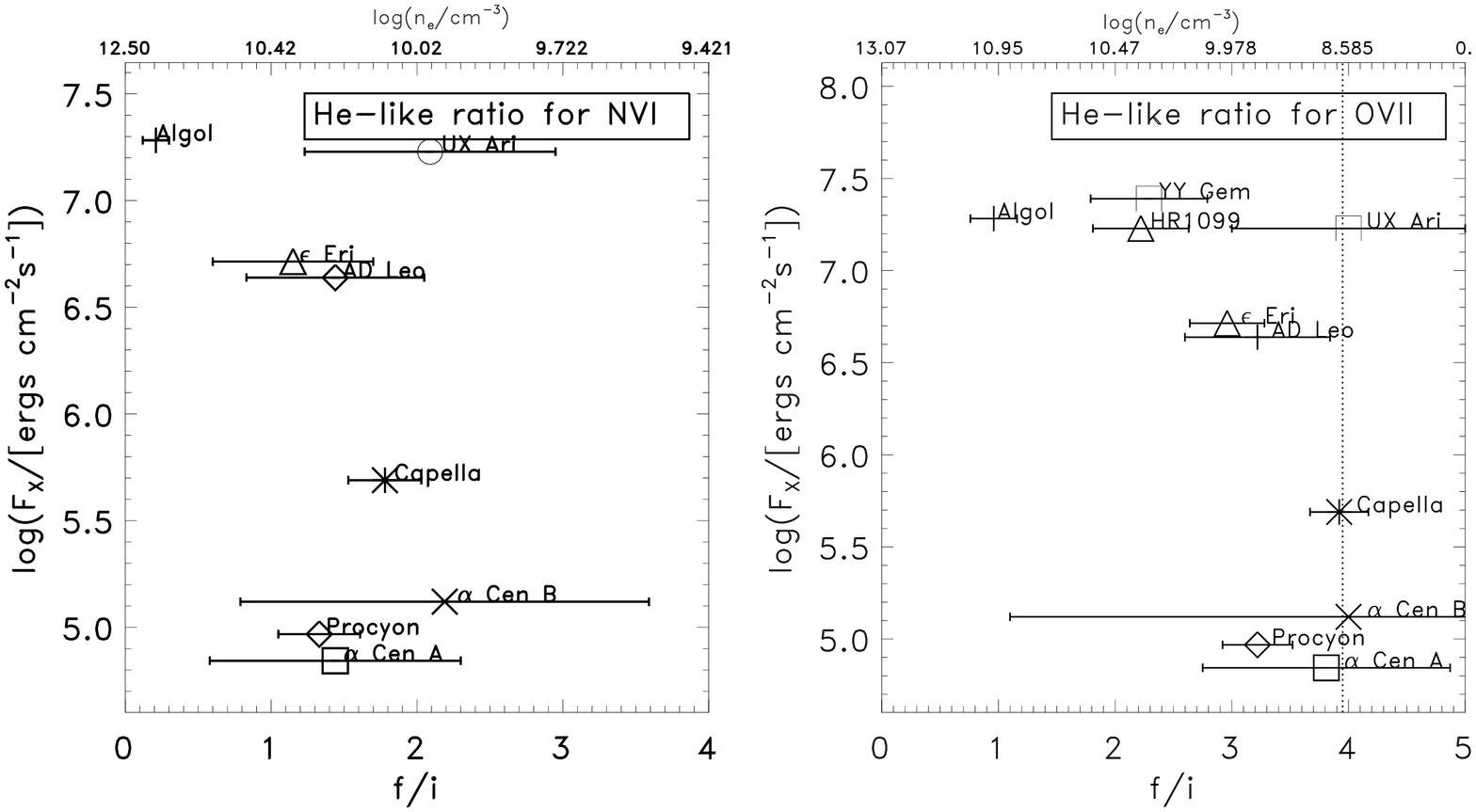}}}
\vspace{-2cm}
\caption[]{\label{rat}f/i ratios for N\,{\sc vi} (left panel) and for
O\,{\sc vii} (left panel) for the stars vs. X-ray surface flux
(values listed in Table~\ref{sprop}). 1\,$\sigma$ errors are given for f/i.
On top of the graph are densities given assuming $\phi/\phi_c=0$ and using the
atomic parameters listed in Table~\ref{restab2}.}

\end{figure*}

In Fig.~\ref{rat} we plot the f/i-ratios versus the X-ray surface flux $F_X$ for
N\,{\sc vi} and O\,{\sc vii}. For a graphical comparison between the densities
in the stars we use the measured f/i ratios instead of densities, because the
conversion of f/i into densities requires atomic data, which might be ambiguous,
while the f/i ratios represent the original measurements. In order to give an
impression of the
corresponding densities we give the $n_e$ scale on top of the graphs using the
atomic data listed in Tables~\ref{restab1} to \ref{restab3}. Unfortunately
the errors on f/i and hence density are still quite large, and for a number
of stars, most notably $\alpha$~Cen A and B, no sensitive statements can be made.
Nevertheless there appears to be a trend that smaller f/i-ratios (and hence high
density) go together with higher density and (from Fig. \ref{LyLx}) with higher
temperature. We note in particular that UX~Ari may in fact have a rather
high density, and the density for Algol may in fact be too high since radiation
effects are bound to play some role unless a very specific viewing geometry
is assumed. It would be extremely interesting to construct diagrams such as
shown in Fig.~\ref{rat} for Ne, Mg, and Si, but this is not possible from the
available LETGS data.

\begin{table}[!ht]
\caption[ ]{\label{restab1}Results for O\,{\sc vii}, N\,{\sc vi}, and C\,{\sc v}
for the ''cooler'' stars Procyon and $\alpha$~Cen A and B.
The errors are 1\,$\sigma$ errors. For each ion the measured He-like ratios
R=f/i and G=(f+i)/r, and the ratios Ly$_\alpha$/r and
Ly$_\alpha/$He=Ly$_\alpha$/(r+i+f) are given. The derived T(Ly$_\alpha$/r),
T(Ly$_\alpha/$He), and T(G) and the densities log($n_e$(f/i)) are also given.
$^1$log($n_e$) denotes the densities accounting for $\phi/\phi_c$;
the radiation terms $\phi/\phi_c$ used for the density
diagnostics were measured with IUE as described by, e.g., Ness et al.
(\cite{ness_cap}). Effective areas are taken from Deron Pease (Oct. 2000).}
\renewcommand{\arraystretch}{1.2}
{\scriptsize
\begin{tabular}{l|ccc}
\hline
&Procyon&\multicolumn{2}{c}{$\alpha$~Cen}\\
&&A&B\\
\hline
HD&61421&128620&128621\\
t$_{exp}$/ksec&140.7&\multicolumn{2}{c}{81.5}\\
\hline
% --------------- O VII -------------------------

\multicolumn{2}{l}{$N_c$=3.4\,10$^{10}$\,cm$^{-3}$,\ $R_0=3.95$}&{\bf O\,{\sc vii}}\\
$T_m$=2.0\,MK&\multicolumn{2}{c}{$^1$ densities accounting for $\phi/\phi_c$}\\
\hline
f/i& 3.22 $\pm$ 0.30&3.81 $\pm$ 1.06&5.6 $\pm$ 2.9\\
(f+i)/r& 1.21 $\pm$ 0.08&1.0 $\pm$ 0.17&1.17 $\pm$ 0.17\\
Ly$_\alpha$/r& 0.67 $\pm$ 0.04 & 0.38 $\pm$ 0.07 & 0.64 $\pm$ 0.08\\
Ly$_\alpha$/He& 0.31 $\pm$ 0.03 & 0.19 $\pm$ 0.06 & 0.30 $\pm$ 0.08\\
$T_{\rm rad}$/K&5410&4653&4181\\
$\phi/\phi_c$&0.01 &0.01&0\\
$^1$log($n_e$/cm$^{-3}$)&9.28 $\pm$ 0.4 &8.96 $\pm$ 1.20&--\\
log($n_e$/cm$^{-3}$)& 9.89 $\pm$ 0.26 & 9.70 $\pm$ 0.58 & --\\
T(Ly$_\alpha$/r)/MK& 2.20 $\pm$ 0.03 & 1.93 $\pm$ 0.08 & 2.17 $\pm$ 0.07\\
T(Ly$_\alpha/$He)/MK& 2.85 $\pm$ 0.04 & 1.68 $\pm$ 0.10 & 1.83 $\pm$ 0.10\\
T(G)/MK&0.96 $\pm$ 0.25&1.67 $\pm$ 0.74&1.09 $\pm$ 0.56\\
\hline

% --------------- N VI -------------------------
\hline
\multicolumn{2}{l}{$N_c$=5.27\,10$^9$\,cm$^{-3}$,\ $R_0=6.0$}&{\bf N\,{\sc vi}}\\
$T_m$=1.4\,MK&\multicolumn{2}{c}{$^1$ densities accounting for $\phi/\phi_c$}\\
\hline
f/i& 1.33 $\pm$ 0.28 &2.19 $\pm$ 0.86&2.09 $\pm$ 1.4\\
(f+i)/r& 0.93 $\pm$ 0.16&1.43 $\pm$ 0.45&1.39 $\pm$ 0.66\\
Ly$_\alpha$/r& 0.94 $\pm$ 0.11 & 0.49 $\pm$ 0.17 & 1.45 $\pm$ 0.49\\
Ly$_\alpha$/He& 0.49 $\pm$ 0.12 & 0.20 $\pm$ 0.10 & 0.61 $\pm$ 0.47\\
$T_{\rm rad}$/K&5780&4626&3994\\
$\phi/\phi_c$&1.58 $\pm$ 0.84 &0.07&0.005\\
$^1$log($n_e$/cm$^{-3}$)& 9.96 $\pm$ 0.23 & 9.95 $\pm$ 0.30&9.99 $\pm$ 0.65\\
log($n_e$/cm$^{-3}$)& 10.27 $\pm$ 0.12 & 9.96 $\pm$ 0.30&10.00 $\pm$ 0.64\\
T(Ly$_\alpha$/r)/MK& 1.09 $\pm$ 0.13 & 1.35 $\pm$ 0.09 & 1.79 $\pm$ 0.16\\
T(Ly$_\alpha/$He)/MK& 1.41 $\pm$ 0.07 & 1.15 $\pm$ 0.09 & 1.43 $\pm$ 0.22\\
T(G)/MK&1.28 $\pm$ 0.46& $<$ 1.2&$<$1.8\\
\hline

% --------------- C V -------------------------

\hline
\multicolumn{2}{l}{$N_c$=6.7\,10$^8$\,cm$^{-3}$,\ $R_0=11.6$}&{\bf C\,{\sc v}}\\
$T_m$=1.0\,MK&\multicolumn{2}{c}{$^1$ densities accounting for $\phi/\phi_c$}\\
\hline
f/i&0.52 $\pm$ 0.13 &2.07 $\pm$ 2.&3.24 $\pm$ 2.86\\
(f+i)/r&1.39 $\pm$ 0.24&0.75 $\pm$ 0.48&1.56 $\pm$ 0.85\\
Ly$_\alpha$/r& 1.78 $\pm$ 0.17 & 1.56 $\pm$ 0.37 & 2.01 $\pm$ 0.60\\
Ly$_\alpha$/He& 0.75 $\pm$ 0.19 & 0.89 $\pm$ 0.89 & 0.79 $\pm$ 0.73\\
$T_{\rm rad}$/K&5532&4864&4124\\
$\phi/\phi_c$&26.67 $\pm$ 9.3 &5.7&0.6\\
$^1$log($n_e$/cm$^{-3}$)& $<$8.92 &$<$ 11.04&9.13 $\pm$ 1.16\\
log($n_e$/cm$^{-3}$)& 10.16 $\pm$ 0.12 & 9.50 $\pm$ 1.56 & 9.24 $\pm$ 1.05\\
T(Ly$_\alpha$/r)/MK& 1.18 $\pm$ 0.03 & 1.14 $\pm$ 0.07 & 1.22 $\pm$ 0.10\\
T(Ly$_\alpha/$He)/MK& 0.95 $\pm$ 0.54 & 1.00 $\pm$ 0.86 & 0.96 $\pm$ 0.17\\
T(G)/MK&$<$0.4& 1.17 $\pm$ 0.8&$<$1.3\\
\hline
\end{tabular}
}
\begin{flushleft}
\renewcommand{\arraystretch}{1}
\end{flushleft}
\end{table}

\begin{table*}
\caption[ ]{\label{restab2}Results for O\,{\sc vii}, N\,{\sc vi}, and C\,{\sc v}
for the ''hotter'' stars corresponding to Table~\ref{restab1}.}
\renewcommand{\arraystretch}{1.2}
{\scriptsize
\begin{tabular}{l|ccccccc}
\hline
&Algol&Capella&$\epsilon$~Eri& UX~Ari&AD~Leo&YY~Gem&HR\,1099\\
\hline
HD&19356&34029&22049&21242&GL388&60179C&22468\\
t$_{exp}$/ksec&81.4&218.5&105.3&112.76&48.5&59&97.5\\
\hline
% --------------- O VII -------------------------
\multicolumn{2}{l}{$N_c$=3.4\,10$^{10}$\,cm$^{-3}$,\ $R_0=3.95$}&\multicolumn{6}{c}{\bf O\,{\sc vii}}\\
$T_m$=2.0\,MK&&&\multicolumn{3}{c}{$^1$ densities accounting for $\phi/\phi_c$}&&\\
\hline
f/i& 0.96 $\pm$ 0.2 &3.92 $\pm$ 0.25 &2.96 $\pm$ 0.32
        &$>$4&3.22 $\pm$ 0.62&2.29 $\pm$ 0.5&2.22 $\pm$ 0.41\\
(f+i)/r& 0.98 $\pm$ 0.17&0.9 $\pm$ 0.03 &0.86 $\pm$ 0.06
        &0.7 $\pm$ 0.14&0.86 $\pm$ 0.11&0.88 $\pm$ 0.14&0.80 $\pm$ 0.10\\
Ly$_\alpha$/r&8.03 $\pm$ 0.71 & 3.49 $\pm$ 0.07 & 2.12 $\pm$ 0.10 &
10.05 $\pm$ 0.93&3.43 $\pm$ 0.1&3.67 $\pm$ 0.32&8.67 $\pm$ 0.53\\
%Ly$_\alpha$/He&4.11 $\pm$ 0.93 & 1.87 $\pm$ 0.12 & 1.14 $\pm$ 0.13 &
%5.94 $\pm$ 10.57&1.86 $\pm$ 0.38&1.97 $\pm$ 0.46&4.86 $\pm$ 0.95\\
$T_{\rm rad}$/K&12710&5030&0&0&0&0&0\\
$\phi/\phi_c$&2.18 $\pm$ 0.29 &0.003 &
        0&0&0&0&0\\
$^1$log($n_e$/cm$^{-3}$)& 10.5 $\pm$ 0.62&$<$9.38 &10.03 $\pm$ 0.23
        &--& 9.89 $\pm$ 0.9&10.39 $\pm$ 0.24&10.42 $\pm$ 0.19\\
log($n_e$/cm$^{-3}$)& 11.04 $\pm$ 0.12 & 8.36 $\pm$ 1.04 &
        9.10 $\pm$ 1.08 & -- &9.89 $\pm$ 0.9&10.39 $\pm$ 0.24&10.42 $\pm$ 0.19\\
T(Ly$_\alpha$/r)/MK&4.81 $\pm$ 0.18 & 3.54 $\pm$ 0.02 &
        3.00 $\pm$ 0.04 & 5.27 $\pm$ 0.20&3.52 $\pm$ 0.08&3.6 $\pm$ 0.1&4.98 $\pm$ 0.12\\
%T(Ly$_\alpha/$He)/MK&4.11 $\pm$ 0.29 & 2.89 $\pm$ 0.05 &
%        2.52 $\pm$ 0.07 & 4.28 $\pm$ 4.28&2.89 $\pm$ 0.17&2.94 $\pm$ 0.21&3.98 $\pm$ 0.26\\
T(G)/MK&1.72 $\pm$ 0.74&2.0 $\pm$ 0.13&2.22 $\pm$ 0.30&3.14 $\pm$ 0.83&2.2 $\pm$ 0.6&2.15 $\pm$ 0.71&2.52 $\pm$ 0.64\\
\hline
% --------------- N VI -------------------------

\multicolumn{2}{l}{$N_c$=5.2\,10$^9$\,cm$^{-3}$,\ $R_0=6.0$}&\multicolumn{6}{c}{\bf N\,{\sc vi}}\\
$T_m$=1.4\,MK&&&\multicolumn{3}{c}{$^1$ densities accounting for $\phi/\phi_c$}&&\\
\hline
f/i& 0.21 $\pm$ 0.09&1.78 $\pm$ 0.25&1.15 $\pm$ 0.55
        &1.44 $\pm$ 0.86&1.07 $\pm$ 0.61&--&--\\
(f+i)/r& 1.66 $\pm$ 0.39&1.33 $\pm$ 0.15 &1.00 $\pm$ 0.39
        &0.69 $\pm$ 0.31&1.48 $\pm$ 0.73&--&--\\
Ly$_\alpha$/r&8.37 $\pm$ 1.28 & 4.91 $\pm$ 0.34 & 3.78 $\pm$ 0.79 &
6.68 $\pm$ 1.10 & 5.48 $\pm$ 1.62&--&--\\
Ly$_\alpha$/He&3.19 $\pm$ 1.38 & 2.14 $\pm$ 0.33 & 1.91 $\pm$ 0.99 &
3.92 $\pm$ 2.44 &2.23 $\pm$ 1.44&--&--\\
$T_{\rm rad}$/K&12060&4980&0&0&0&0&0\\
$\phi/\phi_c$&24.74 $\pm$ 2.41 &0.2 $\pm$ 0.1 &
        0&0&0&0&0\\
$^1$log($n_e$/cm$^{-3}$)& 10.20 $\pm$ 0.93&9.86 $\pm$ 0.12 &
        10.35 $\pm$ 0.33 &10.23 $\pm$ 0.47&10.39 $\pm$ 0.42&--&--\\
log($n_e$/cm$^{-3}$)&11.16 $\pm$ 0.23 & 10.09 $\pm$ 0.09 &
        10.35 $\pm$ 0.32& 10.23 $\pm$ 0.47 &10.39 $\pm$ 0.42&--&--\\
T(Ly$_\alpha$/r)/MK&3.42 $\pm$ 0.23 & 2.73 $\pm$ 0.08 &
        2.47 $\pm$ 0.18 & 3.08 $\pm$ 0.21&2.86 $\pm$ 0.33&--&--\\
T(Ly$_\alpha/$He)/MK&2.32 $\pm$ 0.34 & 2.02 $\pm$ 0.10 &
        1.95 $\pm$ 0.30 & 2.51 $\pm$ 0.53 &2.05 $\pm$ 0.4&--&--\\
T(G)/MK&$<$0.7&0.46 $\pm$ 0.28&$<$ 2.4& 2.1 $\pm$ 0.9&$<$2&--&--\\
\hline

% --------------- C V -------------------------
\hline
\multicolumn{2}{l}{$N_c$=6.7\,10$^8$\,cm$^{-3}$,\ $R_0=11.6$}&\multicolumn{6}{c}{\bf C\,{\sc v}}\\
$T_m$=1.0\,MK&&&\multicolumn{3}{c}{$^1$ densities accounting for $\phi/\phi_c$}&&\\
\hline
f/i& -- &1.60 $\pm$ 0.37 &{\it contaminated}
        &{\it contaminated}&--&--&--\\
(f+i)/r& --&0.91 $\pm$ 0.15 &--&--&--&--&--\\
Ly$_\alpha$/r&-- & 2.55 $\pm$ 0.17 & 2.93 $\pm$ 0.58 &2.87 $\pm$ 0.50
&3.34 $\pm$ 0.93&--&--\\
%Ly$_\alpha$/He&-- & 1.34 $\pm$ 0.32 & & & &--&--\\
$T_{\rm rad}$/K&--&4585&--&--&--&--&--\\
$\phi/\phi_c$&-- &2.54 $\pm$ 0.86&--&--&--&--&--\\
$^1$log($n_e$/cm$^{-3}$)& --&9.42 $\pm$ 0.21 &--&--&--&--&--\\
log($n_e$/cm$^{-3}$)&-- & 9.63 $\pm$ 0.12 &--&--&--&--&--\\
T(Ly$_\alpha$/r)/MK&-- & 1.32 $\pm$ 0.03 & 1.38 $\pm$ 0.08 &1.37 $\pm$ 0.08
        &1.44 $\pm$ 0.13&--&--\\
%T(Ly$_\alpha/$He)/MK&-- & 1.10 $\pm$ 0.06 &--&--&--&--&--\\
T(G)/MK&--&0.85 $\pm$ 0.3&--&--&--&--&--\\
\hline

% --------------- end -------------------------

\end{tabular}
}
\begin{flushleft}
\renewcommand{\arraystretch}{1}
\end{flushleft}
\end{table*}

\begin{table*}
\caption[ ]{\label{restab3}Results for Si\,{\sc xiii}, Mg\,{\sc xi}, and
Ne\,{\sc ix} corresponding to Table~\ref{restab1}. The errors are 1\,$\sigma$
errors including correlated errors in line blends.}
\renewcommand{\arraystretch}{1.2}
{\scriptsize
\begin{tabular}{l|cccccccc}
\hline
&Algol&Capella&$\epsilon$~Eri& UX~Ari&AD~Leo&YY~Gem&HR\,1099&Procyon\\
\hline
HD&19356&34029&22049&21242&Gl388&60179C&22468&61412\\
t$_{exp}$/ksec&81.4&218.5&105.3&112.76&48.5&59&97.5&140.7\\
\hline
% --------------- Si XIII -------------------------
\multicolumn{2}{l}{$N_c$=3.9\,10$^{13}$\,cm$^{-3}$,\ $R_0=2.67$}&\multicolumn{7}{c}{\bf Si\,{\sc xiii}}\\
$T_m$=10.0\,MK&&&&&&&&\\
\hline
f/i& 3.66 $\pm$ 1.45 & 4.41 $\pm$ 0.91& --
        & 3.02 $\pm$ 2.54&1.32 $\pm$ 0.55&--&2.91 $\pm$ 1.17&--\\
(f+i)/r& 0.83 $\pm$ 0.15& 0.67 $\pm$ 0.05 & 0.82 $\pm$ 0.36&0.61 $\pm$ 0.22&0.99 $\pm$ 0.32&--&0.78 $\pm$ 0.16&--\\
Ly$_\alpha$/r&1.54 $\pm$ 0.14 & 0.31 $\pm$ 0.02 & 0.21 $\pm$ 0.14& 1.11 $\pm$ 0.16&0.72 $\pm$ 0.17&0.99 $\pm$ 0.22&1.44 $\pm$ 0.14&--\\
Ly$_\alpha$/He&0.84 $\pm$ 0.34 & 0.18 $\pm$ 0.04 & -- &
0.69 $\pm$ 0.59 &0.36 $\pm$ 0.17&--&0.82 $\pm$ 0.34&--\\
log($n_e$/cm$^{-3}$)& $<$12.9 & $<$12.1
        & $<$14.25 & $<$14.25 &13.6 $\pm$ 0.4&--&$<$13.32&--\\
T(Ly$_\alpha$/r)/MK&14.61 $\pm$ 0.49 & 8.80 $\pm$ 0.14 &
        8.0 $\pm$ 1.1& 12.96 $\pm$ 0.6&11.26 $\pm$ 0.77&12.5 $\pm$ 0.9&14.28 $\pm$ 0.48&--\\
T(Ly$_\alpha/$He)/MK&11.80 $\pm$ 1.41 & 7.75 $\pm$ 0.37 &
        -- & 11.1 $\pm$ 2.54 &9.18 $\pm$ 1.09&--&11.68 $\pm$ 1.46&--\\
T(G)/MK&6.2 $\pm$ 3.0&--&--&11.4 $\pm$ 5.3&$<$9&--&7.1 $\pm$ 4.1&--\\
\hline
% --------------- Mg XI -------------------------

\multicolumn{2}{l}{$N_c$=6.2\,10$^{12}$\,cm$^{-3}$,\ $R_0=2.6$}&\multicolumn{7}{c}{\bf Mg\,{\sc xi}}\\
$T_m$=6.3\,MK&&&&&&&&\\
\hline
f/i& -- & 3.04 $\pm$ 0.53&1.59 $\pm$ 0.63
        & --&--&--&1.72 $\pm$ 0.50&--\\
(f+i)/r& -- & 0.74 $\pm$ 0.04 & 1.06 $\pm$ 0.33 &--&--&--&0.81 $\pm$ 0.16&--\\
Ly$_\alpha$/r& 2.4 $\pm$ 0.31 & 0.59 $\pm$ 0.02 & 0.61 $\pm$ 0.17& 3.24 $\pm$ 1.27&--&1.46 $\pm$ 0.55&1.18 $\pm$ 0.13&--\\
Ly$_\alpha$/He&1.40 $\pm$ 0.60 & 0.34 $\pm$ 0.03 & 0.30 $\pm$ 0.14 &
-- &--&--&0.65 $\pm$ 0.2&--\\
log($n_e$/cm$^{-3}$)& -- & $<$11.35 & 12.6 $\pm$ 0.6&--&--&--&12.5 $\pm$ 0.5&--\\
T(Ly$_\alpha$/r)/MK& 10.43 $\pm$ 0.51 & 6.63 $\pm$ 0.07 & 6.69 $\pm$ 0.49 & 11.7 $\pm$ 1.7 &--&8.74 $\pm$ 1.03&8.14 $\pm$ 0.28&--\\
T(Ly$_\alpha/$He)/MK&8.62 $\pm$ 1.14 & 5.76 $\pm$ 0.12 &
        5.60 $\pm$ 0.55 & -- &--&--&6.82 $\pm$ 0.55&--\\
T(G)/MK& -- &5.8 $\pm$ 0.6&--&--&--&--&4.6 $\pm$ 2.96&--\\
\hline
% --------------- Ne IX -------------------------

\multicolumn{2}{l}{$N_c$=5.9\,10$^{11}$\,cm$^{-3}$,\ $R_0=3.5$}&\multicolumn{7}{c}{\bf Ne\,{\sc ix}}\\
$T_m$=10.0\,MK&&&&&&&&\\
\hline
f/i& 4.31 $\pm$ 1.69 & 2.08 $\pm$ 0.14 & 2.97 $\pm$ 0.49
        & 2.47 $\pm$ 0.37&5.38 $\pm$ 2.13&4.47 $\pm$ 1.79&3.79 $\pm$ 0.7&2.58 $\pm$ 1.41\\
(f+i)/r& 0.68 $\pm$ 0.11& 0.86 $\pm$ 0.04 & 0.82 $\pm$ 0.08
        & 0.75 $\pm$ 0.07&0.71 $\pm$ 0.11&0.70 $\pm$ 0.12&0.74 $\pm$ 0.07&0.72 $\pm$ 0.25\\
Ly$_\alpha$/r&4.56 $\pm$ 0.28 & 1.90 $\pm$ 0.05 & 0.99 $\pm$ 0.07 &
4.04 $\pm$ 0.18 & 1.71 $\pm$ 0.15&2.0 $\pm$ 0.19&4.17 $\pm$ 0.17&0.19 $\pm$ 0.09\\
Ly$_\alpha$/He&2.74 $\pm$ 1.09 & 1.03 $\pm$ 0.07 & 0.55 $\pm$ 0.10 &
2.33 $\pm$ 0.37 & 1.01 $\pm$ 0.41&1.19 $\pm$ 0.49&2.41 $\pm$ 0.46&0.11 $\pm$ 0.08\\
$\phi/\phi_c$&0.1 $\pm$ 0.03 &0 &0 & 0&0&0&0&0\\
log($n_e$/cm$^{-3}$)& $<$11.3& 11.61 $\pm$ 0.07 &11.02 $\pm$ 1.15
        & 11.39 $\pm$ 0.26&$<$10.65&$<$11.25&$<$10.91&11.32 $\pm$ 0.75\\
T(Ly$_\alpha$/r)/MK&7.66 $\pm$ 0.21 & 5.46 $\pm$ 0.05 & 4.44 $\pm$ 0.09
        & 7.28 $\pm$ 0.13 &5.27 $\pm$ 0.16&5.58 $\pm$ 0.17&7.37 $\pm$ 0.12&2.95 $\pm$ 0.26\\
T(Ly$_\alpha/$He)/MK&6.26 $\pm$ 0.88 & 4.49 $\pm$ 0.08 & 3.67 $\pm$ 0.18
        & 5.87 $\pm$ 0.34 &4.47 $\pm$ 0.49&4.68 $\pm$ 0.55&5.94 $\pm$ 0.42&2.67 $\pm$ 0.28\\
T(G)/MK& 4.96 $\pm$ 1.57 & 2.7 $\pm$ 0.4 &3.17 $\pm$ 0.96&4.1 $\pm$ 0.9&4.5 $\pm$ 1.6&4.7 $\pm$ 1.8&4.09 $\pm$ 0.85&4.3 $\pm$ 4.0\\
\hline

\end{tabular}
}
\begin{flushleft}
\renewcommand{\arraystretch}{1}
\end{flushleft}
\end{table*}

\section{Conclusions}
\label{conclusio}

With the data from the new generation of X-ray telescopes we can measure He-like
triplets for a wide range of temperatures and for other stars than the Sun. In
our conclusions we consider only the cases as significant where deviations from
the low-density limit are at least 2\,$\sigma$. We point out that in all our
stars the intercombination line is the weakest line (except for Algol), such
that under exposed spectra suffer the most from large errors in the
intercombination line leading to large uncertainties in the f/i ratios.

The theory of He-like triplets makes specific predictions for the ratio (f+i)/r,
which for collisional dominated plasmas is near unity with some temperature
dependence and for the density-dependent f/i-ratio. For the ``hot'' ions (neon,
magnesium, silicon) the measured (f+i)/r-ratios are near but below unity, the
same applies for the ``cool'' ions (carbon, nitrogen, and oxygen) for the
``hot'', i.e., active stars. For the less active stars (i.e.,
Procyon, $\alpha$ Cen A and B) the (f+i)/r-ratios are somewhat larger,
so we conclude that our measurements are consistent with theoretical
expectations, but unfortunately the accuracy of the measurements is such that
temperature determinations with the help of the (f+i)/r-ratios are in many
cases not possible.

As to densities, an inspection of Table~\ref{restab3} shows that for the
elements neon, magnesium, and
silicon no particularly convincing examples of density sensitive
line ratios were found. Most measurements are consistent with the
low-density limit and with the exception of Capella all ``deviations''
from the low-density limit are found only in one element at a significance
between two and three $\sigma$. We thus conclude that for none of
our sample stars the magnesium and silicon He-like triplets deviate
from the low-density limit. For Capella we derive a neon f/i-ratio of
2.08 $\pm$ 0.14, significantly below the low-density limit. Interestingly,
this value is confirmed by higher resolution measurements with the
{\it Chandra} HEG (cf. Ness et al.~\cite{nessb02}).

Clearly, the best (in terms of signal-to-noise and spectral
resolution) measurements are available for oxygen. For all ten stars the
He-like resonance and forbidden lines were detected. The intercombination
line, which is the weakest triplet line in a plasma with low to intermediate
density, was detected for nine stars. In four cases (i.e., Algol,
$\epsilon$~Eri, YY~Gem, and HR\,1099) do we find significant deviations from
the expected low-density limit, in one case (Procyon) the measured f/i-ratio
is within 2-3 $\sigma$ from the low density limit.
For the stars $\epsilon$~Eri, YY~Gem, and HR\,1099 we find f/i-ratios
between 2.22 - 2.96, but the errors are so large that in principle those
stars could have the same f/i-ratio. The by far lowest f/i-ratio is found for
Algol, but in that particular case radiation fields can be important even for
the oxygen triplet lines. In four cases ($\alpha$\,Cen~A and B, UX~Ari, and
AD~Leo) the measurement errors are so
large that no meaningful conclusions can be drawn, while in one
case (Capella) our measurement agrees perfectly with the low-density
limit. This low-density for Capella is consistent with the
XMM-Newton RGS results, but inconsistent with the measurements obtained
with the {\it Chandra} MEG (cf., e.g., Canizares et al.~\cite{caniz00},
Ayres et al.~\cite{ayres01}, Phillips et al.~\cite{phil01}).

For our analysis we attribute all measured line counts to the expected He-like
lines, which may be contaminated by coincidental lines or by dielectronic
satellite lines. Such effects can be checked in each individual case.
Specfically in the case of Procyon inspection of the MEKAL tables (Mewe,
Kaastra, \& Liedahl \cite{mewe95}) shows that a contribution from dielectronic
satellite lines originating from O\,{\sc vi} to the measured
line flux of the intercombination line of O\,{\sc vii} on the level of 10\%
can easily be accomplished. In this case the measured f/i ratio can well be
consistent with the low-density limit. For the hotter coronae our measurement
errors are larger and further the contribution from dielectronic satellite
lines is weaker. Therefore low f/i ratios cannot be attributed to
contamination of the intercombination line.

What implications can we deduce for the sizes of the underlying coronae?
For the sake of argument, let us focus on the LETGS measurements of YY~Gem.
The available line ratios between Ly$_{\alpha}$ and He-like resonance
lines can be described by a power-law differential emission
measure distribution of the form
$n^2_e \ \frac{dV} {dT} dT = \xi(T) dT = \frac {EM_0} {T_0} (T/T_0)^{\alpha} dT.$
With this functional form we find acceptable fits for the values
$T_0$ = 17\,MK and $\alpha$ = 0. This results in values
of 0.84, 1.69, 2.71, and 3.58 for the line ratios between the Ly$_{\alpha}$
and He-like resonance lines for Si, Mg, Ne, and O, respectively, which compare
well (except for Ne) with our measurements of 0.99 $\pm$ 0.22, 1.46 $\pm$ 0.55,
2.0 $\pm$ 0.19, and 3.67 $\pm$ 0.32, respectively. Calculating the fluxes
of the resonance, forbidden, and intercombination lines of the oxygen triplet
with this emission measure distribution, we find for (f+i)/r a value of
0.89, which must be compared to the measured value of 0.88 $\pm$ 0.14.
The calculated f/i-ratio depends on the assumed pressure. If we assume
a constant pressure scenario, we find an almost linear dependence of pressure
and f/i-ratio, with the measured value of 2.29 corresponding to
17\,dyne\,cm$^{-2}$. Unfortunately, the large errors in f/i make pressures as
low as 10\,dyne\,cm$^{-2}$ and high as 28\,dyne\,cm$^{-2}$ also possible; at any
rate, we can deduce a pressure of 20 $\pm$ 10\,dyne\,cm$^{-2}$ in the corona(e)
of YY~Gem. We note that this value is independent of the precise emission
measure distribution adopted. Fixing then the normalization $EM_0$ from the
requirement to match the observed He-like r flux in oxygen, we can compute
the pressure dependent total coronal volume from the formula
$V_{tot} = \frac {4 k^2 EM_0} {p^2 T_0^2 (\alpha +3)} $
and find $V_{tot} = 4^{+7}_{-2} \times 10^{32}$\,cm$^{3}$; clearly, because
of the quadratic dependence of the coronal volume on pressure the
errors are still substantial. Nevertheless, the nominal value of
$V_{tot} = 4 \times 10^{32}$\,cm$^{3}$ leads (for an assumed stellar
radius of 0.5 R$_{\odot}$, two identical stars and filling factor of unity)
to a height of $\approx 10^{10}$\,cm, i.e., about one third of the stellar
radius and hence much larger than typical solar coronal loops.

Using the line ratio f/i of the He-like triplets of oxygen and nitrogen one
recognizes a trend, indicating lower f/i-ratios and thus lower densities
for inactive stars. This conclusion has, however, to be treated with some care.
The available data support it only for main sequence stars, i.e., for YY~Gem,
AD~Leo, and $\epsilon$~Eri do we measure f/i-ratios significantly below the
low-density limit. For Algol the measured low f/i-ratio does not necessarily
imply high densities, in principle it can also be attributed to a special
geometrical configuration immersing its corona in the radiation field of
its B-type companion (cf. Sect.~\ref{radfield}). For UX~Ari the data are
consistent with the low density limit, but because of the large errors
significant f/i ratios can also not be ruled out. Only for Capella the
measurement errors are small enough that we can state with confidence
that the measured f/i-ratio places Capella very close to the low-density limit.
For Capella one can definitely exclude densities as high as
10$^{13}$\,cm$^{-3}$, which were reported by Dupree et al. (\cite{dup93})
from EUVE measurements
of Fe\,{\sc xxi} lines, which are, however, formed at a much higher temperature
than O\,{\sc vii}. The analysis of the long wavelength portion of the
{\it Chandra} LETGS spectrum by Mewe et al. (\cite{mewe_cap}) yielded only
upper limits of $\approx$ 2 $\times 10^{12}$\,cm$^{-3}$ for Fe\,{\sc xxi}, and
similarly for their He-like line ratios of Mg\,{\sc xi} and Si\,{\sc xiii} no
significant densities could be derived. The densities derived from
N\,{\sc vi} show the same trend as those derived from oxygen. Again we
emphasize that the peak of the emission measure distribution for the most
active stars lies above the peak formation temperatures of those specific ions.
For the lowest temperature ion, C\,{\sc v}, the f/i-ratios for the hotter
stars are dominated by radiation effects; also higher order line blending
limits the accuracy with which the C\,{\sc v}-triplet can be determined.
In all cases where the C\,{\sc v}-triplet could be detected f/i-ratios
of approximately unity were found. The higher Z ions Ne to Si are more
difficult to study with the LETGS, since they are not fully resolved and
line blending occurs especially for Ne\,{\sc ix}. These line blends can still
be modelled and thus de-blended, but our results should be confirmed with higher
resolution measurements available with, e.g., the MEG. The mean X-ray
surface flux appears to be a useful
parameter segregating stars into different levels of activity; another useful
parameter appears to be the ratio between Ly$_{\alpha}$ line flux and the
He-like triplet resonance line flux, especially for oxygen.
A grouping of stars is apparent when mean X-ray surface flux and
the ratio between Ly$_{\alpha}$ and He-like triplet flux are considered.
Larger data samples are required to assess any physical significance of such
groupings and open the road to an X-ray ``HR-diagram''.

\begin{acknowledgements}
J.-U.N. acknowledges financial support from Deutsches Zentrum f\"ur Luft- und
Raumfahrt e.V. (DLR) under 50OR98010.\\
The Space Research Organization Netherlands (SRON) is supported financially by NWO.
\end{acknowledgements}

\end{document}